\documentclass[twocolumn]{aastex63}

\newcommand{\eps}{\varepsilon}
\newcommand{\xijtrue}{x_{ij}^{\textrm{true}}}

\newcommand{\xijobs}{x_{ij}^{\textrm{obs}}}
\newcommand{\xiobs}{x_{i}^{\textrm{obs}}}
\newcommand{\xijrep}[1]{x_{ij(#1)}^{\textrm{rep}}}
\newcommand{\xirep}[1]{x_{i(#1)}^{\textrm{rep}}}
\newcommand{\xtrue}[1]{x_{#1}^{\textrm{true}}}

\usepackage{amssymb}
\usepackage{graphicx}
\usepackage{array, booktabs}
\usepackage{comment}
\usepackage{multirow}
\usepackage{makecell}
\usepackage{dsfont}
\usepackage{graphbox} % for aligning graphics [align=c]
\usepackage[ruled,vlined]{algorithm2e}

\usepackage{lineno}
\usepackage{amsmath}
\usepackage{etoolbox}

\newcommand*\linenomathpatch[1]{%
  \cspreto{#1}{\linenomath}%
  \cspreto{#1*}{\linenomath}%
  \csappto{end#1}{\endlinenomath}%
  \csappto{end#1*}{\endlinenomath}%
}

\linenomathpatch{equation}
\linenomathpatch{gather}
\linenomathpatch{multline}
\linenomathpatch{align}
\linenomathpatch{alignat}
\linenomathpatch{flalign}

\shorttitle{Gaussian Perturbation for Classification with Measurement Error}
\shortauthors{Shy et al.}
\begin{document}

\title{Incorporating Measurement Error in Astronomical Object Classification}

\correspondingauthor{Hyungsuk Tak}\email{tak@psu.edu}

\author{Sarah Shy}
\affiliation{Department of Statistics, Pennsylvania State University, University Park, PA 16802, USA}

\author[0000-0003-0334-8742]{Hyungsuk Tak}
\affiliation{Department of Statistics, Pennsylvania State University, University Park, PA 16802, USA}
\affiliation{Department of Astronomy and Astrophysics, Pennsylvania State University, University Park, PA 16802, USA}
\affiliation{Center for Astrostatistics, Pennsylvania State University, University Park, PA 16802, USA}
\affiliation{Institute for Computational and Data Sciences, Pennsylvania State University, University Park, PA 16802, USA}

\author{Eric D. Feigelson}
\affiliation{Department of Astronomy and Astrophysics, Pennsylvania State University, University Park, PA 16802, USA}
\affiliation{Department of Statistics, Pennsylvania State University, University Park, PA 16802, USA}
\affiliation{Center for Astrostatistics, Pennsylvania State University, University Park, PA 16802, USA}

\author{John D. Timlin}
\affiliation{Department of Astronomy and Astrophysics, Pennsylvania State University, University Park, PA 16802, USA}

\author{G. Jogesh Babu}
\affiliation{Department of Statistics, Pennsylvania State University, University Park, PA 16802, USA}
\affiliation{Department of Astronomy and Astrophysics, Pennsylvania State University, University Park, PA 16802, USA}
\affiliation{Center for Astrostatistics, Pennsylvania State University, University Park, PA 16802, USA}

\begin{abstract}

Most general-purpose classification methods, such as support-vector machine (SVM) and random forest (RF), fail to account for an unusual characteristic of astronomical data: known measurement error uncertainties. In astronomical data, this information is often given in the data but discarded because popular machine learning classifiers cannot incorporate it. We propose a simulation-based approach that incorporates heteroscedastic measurement error into  existing classification method to better quantify uncertainty in classification. The proposed method first simulates perturbed realizations of the data from a Bayesian posterior predictive distribution of a Gaussian measurement error model. Then, a chosen classifier is fit to each simulation. The variation across the simulations naturally reflects the uncertainty propagated from the measurement errors in both labeled and unlabeled data sets. We demonstrate the use of this approach via two numerical studies. The first is a thorough simulation study applying the proposed procedure to SVM and RF, which are well-known hard and soft classifiers, respectively. The second study is a realistic classification problem of identifying high-$z$ $(2.9 \leq z \leq 5.1)$ quasar candidates from photometric data. The data \textcolor{black}{are} from merged catalogs of the Sloan Digital Sky Survey, the \textit{Spitzer} IRAC Equatorial Survey, and the \textit{Spitzer}-HETDEX Exploratory Large-Area Survey. The proposed approach reveals that out of 11,847 high-$z$ quasar candidates identified by a random forest without incorporating measurement error, 3,146 are potential misclassifications \textcolor{black}{with measurement error}. Additionally, out of $1.85$ million objects not identified as high-$z$ quasars without measurement error, 936  can be considered \textcolor{black}{new} candidates \textcolor{black}{with} measurement error. 

\end{abstract}


%%%%%%%%%%%%%%%%%%%%%%%%%%%%%%%%%%%%%%%%%%%%%%%%%%%%%%%%%%%%%%%%%%%%%%%%%%%%%%%%

%%%%%%%%%%%%%%%%%%%%%%%%%%%%%%%%%%%%%%%%%%%%%%%%%%%%%%%%%%%%%%
\section{Introduction} \label{sec:intro}
%%%%%%%%%%%%%%%%%%%%%%%%%%%%%%%%%%%%%%%%%%%%%%%%%%%%%%%%%%%%%%

\subsection{Motivation}

% history + problem
Classification methods have played an important role in astronomy for centuries and continue to be essential tools in modern astronomy \citep[chap.~9]{feigelson2012classification}. According to the Astrophysics Data System, the keyword `classification' appeared in over 6000 refereed astronomy papers annually for the last 5 consecutive years (2017--2021). One critical methodological issue is that most standard classification methods\textcolor{black}{, originally developed outside astronomy,} do not have built-in functionality to account for measurement error, a notable property of astronomical data. As a result, to implement these classification methods, we have no choice but to assume that the data are error-free measurements. This assumption inevitably produces over-confident outcomes. It is therefore essential to account for measurement error in classification analyses to reliably classify astronomical objects.

\subsection{Past Studies}

\subsubsection{Measurement Error Studies}

% related work
Incorporating measurement error into classification analysis is an active area of research across various scientific fields. However, classical statistical treatments of measurement error are, by and large, limited to regression problems, and homoscedasticity is often assumed \citep{fuller1987measurement, carroll2006measurement, buonaccorsi2010measurement}. The most common approach in machine learning is to use a weighting scheme, applying heavier weights to more certain measurements \citep{hoefsloot2006maximum, lapin2014learning, hashemi2018weighted, luo2019classification}. The inverse of the error variance is often used for the weight, as in the case of classical weighted linear regression. However, these solutions are method-specific, requiring a unique modification to each classification method. Moreover, they consider measurement error in the training set only. Existing method-agnostic approaches to account for heteroscedastic measurement error are limited to scaling and transforming the raw data according to the measurement error variance \citep{waaijenborg2018fusing}. Common transformations include the square root and log transformation \citep{van2006centering}. Such transformations assume a specific, strictly monotonic, relationship between measurement intensity and measurement error variance. A more general approach that applies to any error structure is needed.

In astronomical literature specifically, the gap in statistical methodology for incorporating heteroscedastic measurement error has been noted for many years \citep{feigelson1998statistical}. Various astronomical contributions have been made dating back to \citet{eddington1913formula}, where the influence of small observational error on flux measurements was first investigated. \citet{petrosian1992luminosity}, \citet{bisnovatyi1997gamma}, and \citet{hogg1998maximum} similarly discuss measurement error on univariate measurements. In regression, \citet{akritas1996linear}, \citet{kelly2007some}, \citet{andreon2013measurement}, and \citet{sereno2016bayesian} present notable contributions in both classical regression and Bayesian approaches. In time-domain astronomy, the heteroscedastic measurement error has been properly modeled in univariate and multivariate damped random walk processes \citep{kelly2009variations, hu2020} and the more general univariate CARMA($p, q)$ process \citep{kelly2014flexible}. \textcolor{black}{In classification, \cite{2011ApJ...729..141B, 2012ApJ...749...41B} and \cite{ 2015MNRAS.452.3124D} incorporate the  measurement error into their unique model-based probabilistic classifiers to identify quasar candidates according to redshift.} However, the disconnect between the astronomical need for incorporating measurement error and the existing statistical methods for doing so is still ongoing for many other data analyses \citep{feigelson2021twenty}.

% dealing with various quasar classification problems according to 

\subsubsection{Data Simulation Approaches}

\textcolor{black}{Unlike model-based classification methods \citep[e.g.,][]{2011ApJ...729..141B, 2012ApJ...749...41B, 2015MNRAS.452.3124D}, simulation-based methods have not received much attention in astronomy. However,}  simulation-based approaches, such as bootstrapping  \citep{efron1992bootstrap, efron1994introduction}, subsampling \citep{vonLuxburg2010clustering}, and random projections on a lower dimensional space \citep{achlioptas2003database} are widely used in statistical machine learning \textcolor{black}{for classification}. For example, (i) \citet{yu2013stability} and \citet{sun2015stability} review various ways to simulate data sets through bootstrapping, subsampling, and adding random noise in the context of classification stability and reproducibility. (ii) \citet{cannings2021random} uses random projections to create simulated data sets in lower dimensions and aggregates the results to produce an ensemble classifier with better predictive power. (iii) \citet{malossini2006detecting} detect potential labeling errors by flipping one label at a time and assessing the sensitivity of the results to a single flipped observation.  (iv) \citet{darling2018toward} characterize the distribution of simulated model outputs for decision trees and $k$-nearest neighbors by bootstrapping and sampling-without-replacement. However, these works do not incorporate \textit{known} measurement error information. In addition, measurement error in the unlabeled data (whose labels are to be predicted) is not considered. Thus, these methods are still inadequate for a large array of classification problems in astronomy.

% \begin{figure*}[t!]
%     \fig{gp_pipeline2.pdf}{1\textwidth}{}
%     \caption{The proposed framework uses the original observed data to simulate multiple perturbed data sets. Each perturbed data set is independently drawn from the posterior predictive distribution of a Gaussian measurement error model. We re-fit the classifier to each perturbed data set and calculate any metrics of interest, such as prediction accuracy. This will produce a posterior predictive distribution for each metric, from which we can assess uncertainty by the spread of the distribution. Consequently, the measurement error uncertainty in the original data is propagated through each step in the framework.\label{fig:pipelineDiagram}}
% \end{figure*}

\begin{figure*}[t!]
    \includegraphics[width = \textwidth]{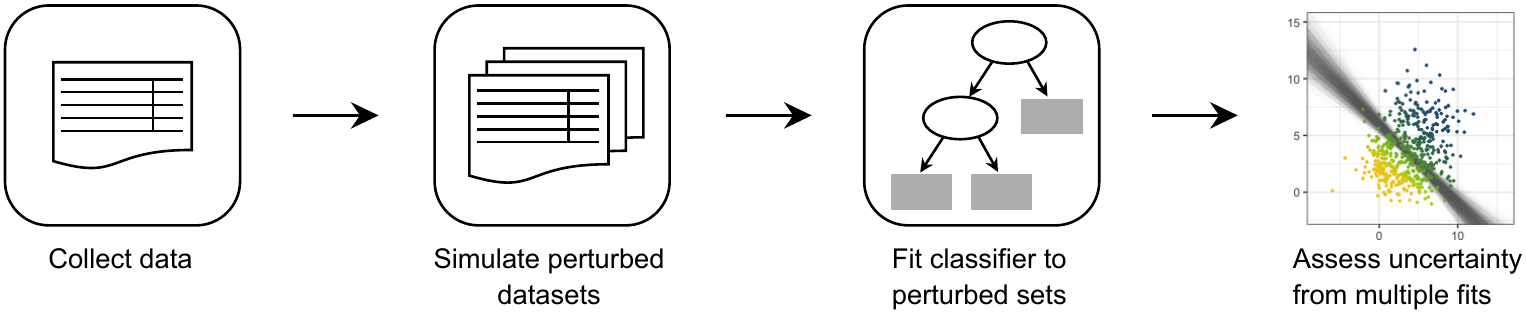}
    \centering
    \caption{The proposed framework uses the original observed data to simulate multiple perturbed data sets. Each perturbed data set is independently drawn from the posterior predictive distribution of a Gaussian measurement error model. We re-fit the classifier to each perturbed data set and calculate any metrics of interest, such as prediction accuracy. This will produce a posterior predictive distribution for each metric, from which we can assess uncertainty by the spread of the distribution. Consequently, the measurement error uncertainty in the original data is propagated through each step in the framework.}
    \label{fig:pipelineDiagram}
\end{figure*}

\subsection{Our Contributions}

% introduce our approach
To overcome the limitations of existing methods, we propose a simulation-based approach that incorporates heteroscedastic measurement error into any \textcolor{black}{existing} classification algorithm. In principle, the proposed \textcolor{black}{method, called Gaussian perturbation,} can quantify the uncertainty of \textit{any} quantity of interest relating to the classification results. First, \textcolor{black}{it} simulate\textcolor{black}{s} multiple pseudo-data sets using the known measurement error uncertainties with minimal assumptions on a Gaussian measurement error model. Second, we fit a classifier on each simulation, using the same method each time. Finally, from the ensemble of classifications, we quantify classification uncertainty by the variation across the \textcolor{black}{multiple} fits. Consequently, \textcolor{black}{this} approach propagates uncertainties of heteroscedastic measurement error from both the labeled and unlabeled datasets to the final classification results. Figure~\ref{fig:pipelineDiagram} illustrates this process. 

We note \textcolor{black}{two important features of the proposed Gaussian perturbation method. First, the proposed method is not a  classification method. Instead, it is a simulation framework like bootstrapping to encode  measurement error uncertainty into any classifiers that are not originally designed to incorporate measurement error. Examples of such classifiers include (but are not limited to) support-vector machine, random forest, or deep learning  neural network. Second,}  the proposed method is not intended to improve classification accuracy. Classification cannot be more accurate after incorporating additional uncertainty than if we ignore measurement error. Instead, the goal is to present more reliable and informative classification results by incorporating uncertainty from measurement error \textcolor{black}{into classifiers that do not naturally account for it}.

\begin{figure}[t!]
  \includegraphics[scale = 1]{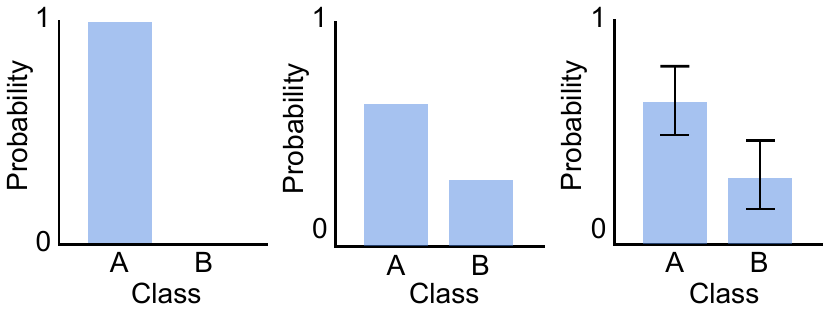}
  \centering
  \caption{Left: For a single observation, a hard classifier outputs a single predicted class label, i.e., the object is predicted to be in that class with probability 1. Center: A soft classifier reports the relative probability of falling into each class. Right: A further softened soft classifier assesses the variability in the class probabilities. The proposed method converts a hard classifier into a soft one (left to center) and further softens soft classifiers (center to right).}
  \label{fig:softenClassifier}
\end{figure}

We illustrate the proposed approach using two popular classification methods: support-vector machine and random forest (hereafter SVM and RF, respectively). The former is known as a hard classifier because the final classification result is a single predicted class label, i.e., one predicted class with probability 1 \citep{wahba2002soft}. The latter is a soft classifier in the sense that for every object, it provides a relative probability (estimate) of belonging to each class. The proposed approach absorbs the known measurement error uncertainties to soften hard classifiers and further soften soft classifiers. Softening hard classifiers means that the information about the probability of belonging to each class, which is naturally given by soft classifiers, becomes available to hard classifiers. Softening soft classifiers means that we are able to assess the variability of the class probabilities provided by soft classifiers. This concept is illustrated in Figure~\ref{fig:softenClassifier}. We note that the applicability of the proposed approach is not restricted to these two specific classifiers but can be applied to any classifier, including deep learning neural networks. \textcolor{black}{In principle, any statistical methods, such as regression, clustering, and time series analysis, can be fit into Gaussian perturbation  to quantify extra uncertainty from astronomical measurement error.}

% \begin{figure}[t!]
%   \fig{soften_classifier.pdf}{1\columnwidth}{}
%   \caption{Left: For a single observation, a hard classifier outputs a single predicted class label, i.e., the object is predicted to be in that class with probability 1. Center: A soft classifier reports the relative probability of falling into each class. Right: A further softened soft classifier assesses the variability in the class probabilities. The proposed method converts a hard classifier into a soft one (left to center) and further softens soft classifiers (center to right).\label{fig:softenClassifier}}
% \end{figure}

To better present the advantages of the proposed approach, we conduct a simulation study and a realistic classification analysis of high-redshift ($2.9 \leq z \leq 5.1$) quasars. Both studies show that classification without considering measurement error produces over-confident results when classifying individual objects into two classes. For example, an object is confidently classified into one class even when its measurement error bar spans across the classification boundary. The proposed approach captures this uncertainty, showing that the object is not classified into one class dominantly across all simulations. In the high-$z$ quasar classification, random forest identifies 11,847 high-$z$ quasar candidates out of ${\sim1.86}$ million objects when measurement error is not considered. On the other hand, the proposed approach reveals that 3,146 out of the 11,847 \textcolor{black}{(26.6\%)} are potential misclassifications when their measurement errors are considered. In addition, of the ${\sim1.85}$ million objects \textcolor{black}{(haystack)} not identified to be high-$z$ quasars without considering measurement error, 936 \textcolor{black}{(needles)} should be considered potential candidates once we account for measurement error. These results are based on a simple decision threshold of 0.5, i.e., we have classified an object as a high-$z$ quasar if its estimated probability of being a high-$z$ quasar were greater than 0.5.

% paper outline
The rest of this article is organized as follows. In Section~\ref{sec:data}, we briefly review the data format and notation for classification. In Section \ref{sec:methodology}, we specify the details of the proposed approach as a generic way to incorporate heteroscedastic measurement error into any standard classification method. Section \ref{sec:simulation} presents a thorough simulation study demonstrating how the proposed procedure can be applied to SVM and RF. In Section \ref{sec:astro}, we present a realistic astronomical application to identify high-$z$ quasar candidates. Finally, we discuss potential limitations of the proposed work and future directions in Sections \ref{sec:discussion} and \ref{sec:conclusion}\textcolor{black}{, respectively.}

%%%%%%%%%%%%%%%%%%%%%%%%%%%%%%%%%%%%%%%%%%%%%%%%%%%%%%%%%%%%%%
\section{Data for classification in astronomy}\label{sec:data}
%%%%%%%%%%%%%%%%%%%%%%%%%%%%%%%%%%%%%%%%%%%%%%%%%%%%%%%%%%%%%%

% \begin{figure}[t!]
%     \centering
%     \begin{subfigure}{0.48\columnwidth}
%         \centering
%         \includegraphics[scale = 0.1]{typical_data_format.png}
%         \caption{ }
%     \end{subfigure}
%     \begin{subfigure}{0.48\columnwidth}
%         \centering
%         \includegraphics[scale = 0.1]{astro_data_format.png}
%         \caption{ }
%     \end{subfigure}
%     \caption{In a typical classification problem, a classifier is trained and validated on the labeled set, and then used to make predictions on the unlabeled set. In astronomy, the information about measurement error uncertainty is additionally given in the data set, as illustrated in the right panel.}    \label{fig:data_split}
% \end{figure}

\begin{figure}[t!]
    \centering
    \gridline{\fig{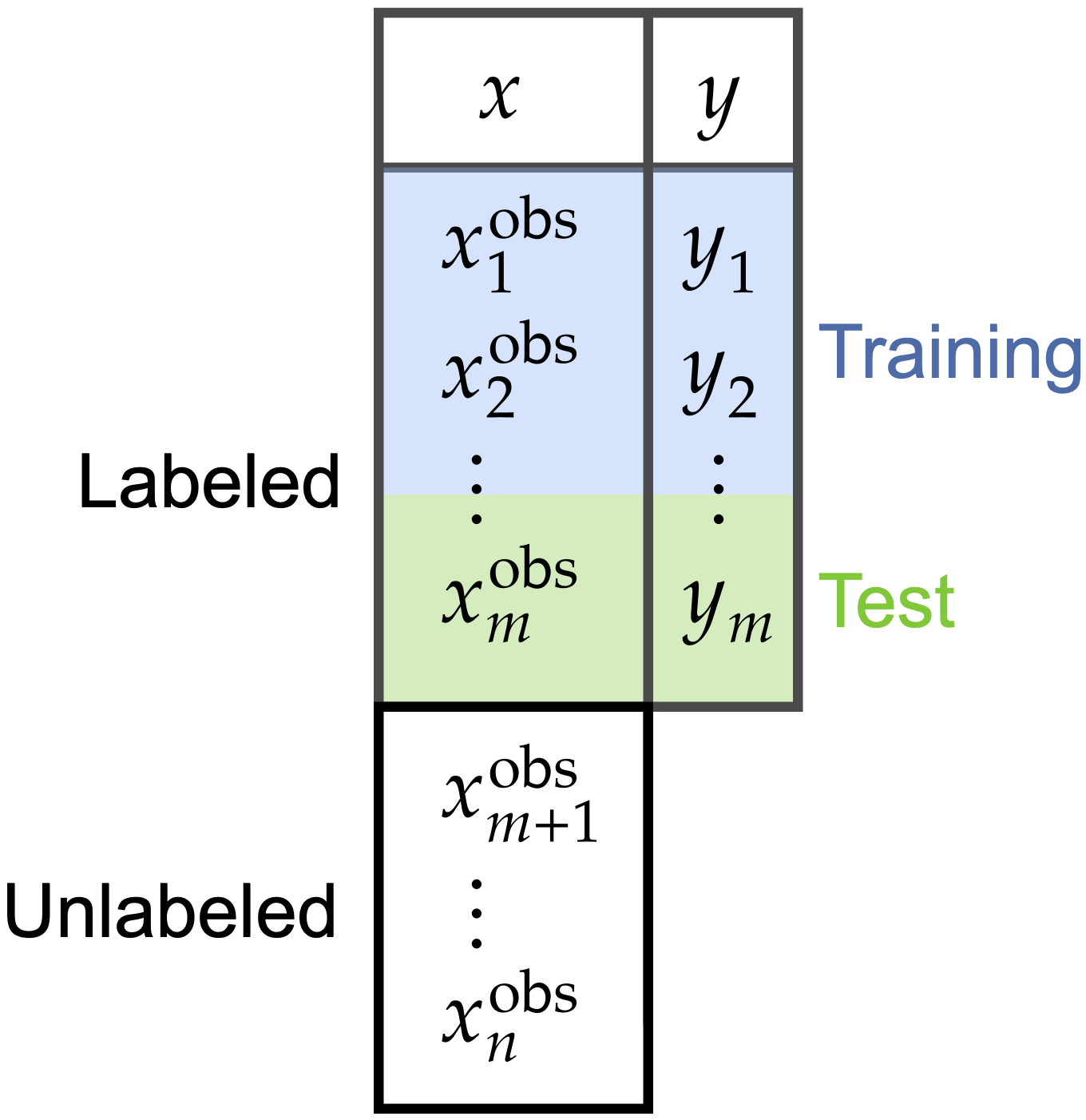}{0.44\columnwidth}{(a)}
              \fig{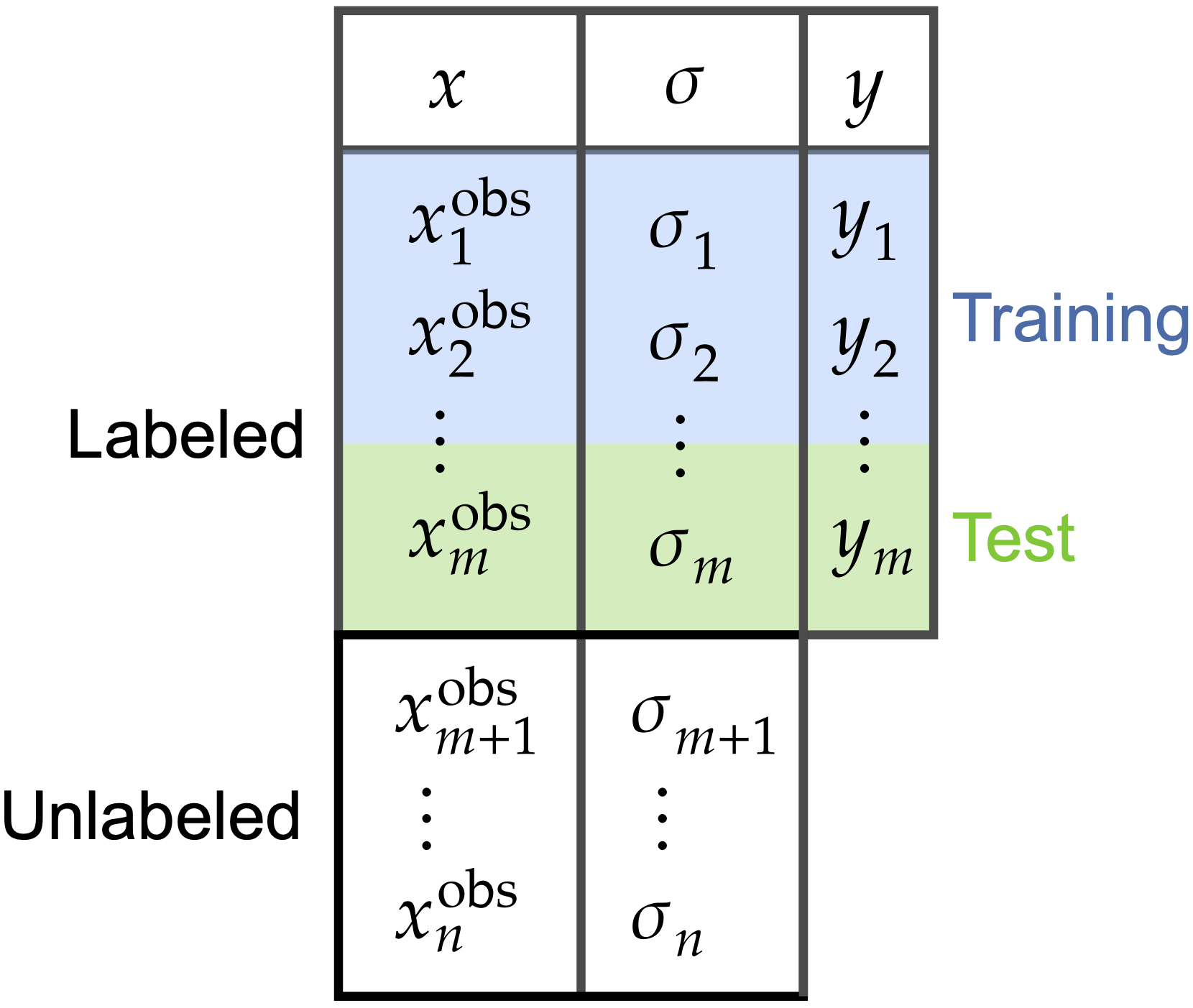}{0.54\columnwidth}{(b)}}
    \caption{In a typical classification problem, a classifier is trained and validated on the labeled set, and then used to make predictions on the unlabeled set. In astronomy, the information about measurement error uncertainty is additionally given in the data set, as illustrated in the right panel.}    \label{fig:data_split}
\end{figure}

In a typical classification problem, the data are composed of labeled and unlabeled datasets, as shown in the left panel of Figure~\ref{fig:data_split}. For convenience, we assume that the entire sample of $n$ objects is unified, with the first $m$ observations being labeled and the remaining unlabeled.  The $i$-th row contains measurements of $p$ features for the $i$-th object, which is denoted by  $\xiobs=\{x^{\textrm{obs}}_{i1}, x^{\textrm{obs}}_{i2}, \ldots, x^{\textrm{obs}}_{ip}\}$. In the labeled set, the class label of the $i$-th object is denoted by $y_i$.  In practice, the labeled set is often randomly split into training and test sets (e.g., in $k$-fold cross validation) to train and validate a classifier. Then, the fitted classifier is used to make predictions on the unlabeled dataset.

In astronomy, a measurement of the $j$-th feature for the $i$-th object (i.e., $\xijobs$) is subject to measurement uncertainty. This is due to limitations of the telescope and detector sensitivity, exposure time, observing conditions, and data reduction procedures. The common assumption is that the measurement errors are i.i.d.~Gaussian noise \citep{eddington1913formula}, i.e.,
\begin{equation*}
\epsilon_{ij}\sim \mathcal{N}(0,~ \sigma^2_{ij})
\end{equation*}
The 1$\sigma$ measurement error uncertainty, $\sigma_{ij}$, is directly measured by careful calibration of the instrument and examination of source-free regions of the image or spectrum. The phenomenon where the values of $\sigma_{ij}$ differ for each object $i$ as well as for each feature $j$ is called \emph{heteroscedasticity}.

Thus, the heteroscedastic measurement error uncertainty is additionally known for both the labeled and unlabeled sets in astronomy, as illustrated in the right panel of Figure~\ref{fig:data_split}. In the figure,  we denote the $1\sigma$ uncertainties of $p$ measurements of the $i$-th object by $\sigma_i=\{\sigma_{i1}, \sigma_{i2}, \ldots, \sigma_{ip}\}$. These measurement error uncertainties are often ignored out of necessity. This work aims at utilizing the entire array of available information, including the  heteroscedastic measurement error uncertainties given in the data.

%%%%%%%%%%%%%%%%%%%%%%%%%%%%%%%%%%%%%%%%%%%%%%%%%%%%%%%%%%%%%%
\section{Methodology} \label{sec:methodology}
%%%%%%%%%%%%%%%%%%%%%%%%%%%%%%%%%%%%%%%%%%%%%%%%%%%%%%%%%%%%%%

We adopt \textcolor{black}{the} Gaussian measurement error model, assuming that the observed quantities are noisy measurements centered at the true quantities with known measurement error uncertainty \citep{eddington1913formula}. This means that hypothetical repeated measurements of the same object will follow a Gaussian distribution centered at its true value. For each of $n$ observations ($i=1, 2, \ldots, n$) and $p$ features ($j=1, 2, \ldots, p$),
\begin{equation}\label{obs_dist}
    \xijobs \sim \mathcal{N}(\xijtrue,~ \sigma_{ij}^2),
\end{equation}
where $\xijobs$ is the measurement of the $j$-th feature of the $i$-th object, $\xijtrue$ is the corresponding unknown true quantity, and $\sigma_{ij}$ is the known  measurement error uncertainty. Here we assume an independent measurement for each feature, but we also discuss correlated measurements in Section~\ref{correlated}.

%, and to minimize the modeling uncertainty
To reflect our lack of knowledge about the true values, $\xijtrue$'s, we let the data speak for themselves by adopting a jointly improper flat prior on $\xijtrue$. That is, for all $i$ and $j$,
\begin{equation}\label{prior1}
h(\xijtrue)\propto 1.
\end{equation}
The resulting posterior distribution of the true value $\xijtrue$ given the observation $\xijobs$ is a proper Gaussian distribution,
\begin{equation} \label{eq:post_dist}
    \xijtrue \mid \xijobs \sim \mathcal N(\xijobs,~ \sigma_{ij}^2),
\end{equation}
satisfying posterior propriety \citep{1996hobert, 2018MNRAS.481..277T}.

To simulate a perturbed replicate of the observed data, we sample the posterior predictive distribution of the Gaussian measurement error model,
\begin{equation*}
q(x_{ij}^{\textrm{rep}} \mid \xijobs)=\int f(x_{ij}^{\textrm{rep}} \mid \xijtrue)~\pi(\xijtrue\mid \xijobs)~d\xijtrue,
\end{equation*}
where $x_{ij}^{\textrm{rep}}$ is a predicted value of $\xijobs$. The distributions $f$ and $\pi$ in the integrand  are defined in Equations \eqref{obs_dist} and \eqref{eq:post_dist}, respectively. The distribution of $(x_{ij}^{\textrm{rep}} \mid \xijtrue)$ is the same as that of $(x_{ij}^{\textrm{obs}} \mid \xijtrue)$ because both $x_{ij}^{\textrm{rep}}$ and $x_{ij}^{\textrm{obs}}$ are noisy measurements given the unknown true value. This posterior predictive distribution is a distribution of  predicted data given the observed data, which we obtain by accounting for all possible realizations of the unknown true value $\xijtrue$ \citep[Section~1.3]{gelman2013bayesian}. The resulting posterior predictive distribution of $q(x_{ij}^{\textrm{rep}} \mid \xijobs)$ is still Gaussian:
\begin{equation} \label{eq:ppd}
x_{ij}^{\textrm{rep}}\mid \xijobs \sim \mathcal{N}(\xijobs,~ 2\sigma_{ij}^2).
\end{equation}
Here, the prediction variance $2\sigma_{ij}^2$ is inflated from the variance of the measurement error $\sigma_{ij}^2$ by a factor of~2.  This is to account for the additional uncertainty of the unknown true value $\xijtrue$ when predicting a future observation $x_{ij}^{\textrm{rep}}$ given the proposed model. In other words, the inflation factor is the result of our choice of prior in~\eqref{prior1} \textcolor{black}{under the Gaussian measurement error model}. For example, if we were to use a different Gaussian prior, the resulting posterior predictive distribution would be still Gaussian, but with a different variance. 
 
This Gaussian posterior predictive distribution specifies that we can directly simulate replicates of $\xijobs$ by generating $x_{ij}^{\textrm{rep}}$ from the Gaussian distribution in Equation~\eqref{eq:ppd}. We call this replication process \emph{Gaussian perturbation} because $x_{ij}^{\textrm{rep}}$ can be considered a perturbation of $\xijobs$ with random noise drawn from $\mathcal{N}(0,~ 2\sigma_{ij}^2)$. Performing this process repeatedly will produce multiple simulations of the observed data under the proposed Gaussian measurement error model. We denote the $r$-th replicate of the $i$-th observation by $\xirep{r}=\{x_{i1(r)}^{\textrm{rep}}, x_{i2(r)}^{\textrm{rep}}, \ldots, x_{ip(r)}^{\textrm{rep}}\}$ and the $r$-th simulated data set by $X_{(r)}^{\textrm{rep}}=\{\xirep{r}: i=1, 2, \ldots, n\}$.  This simulation process is summarized in Algorithm \ref{alg:perturb_data}. 

\begin{algorithm}[b!]
    \SetAlgoLined
    \SetKwInOut{Input}{input}
    \SetKwInOut{Output}{result}

    \Input{Observed data $\xijobs$'s, known measurement error uncertainties $\sigma_{ij}$'s}
    \Output{$R$ perturbed data sets $X_{(1)}^{\mathrm{rep}}, \ldots, X_{(R)}^{\mathrm{rep}}$} 
    \For{$r=1 \dots R$}{
        Sample $\xijrep{r}\sim \mathcal N(\xijobs,~  2\sigma_{ij}^2)~~\textrm{for all}~i~\textrm{and}~j$
        % Fit classifier $C$ using perturbed data $\xirep{s}$ \\
        % Record new predictions $\hat{\mathbf y}^{(b)} = C(\mathbf W_i^{*(b)})$ \\
     }
    \caption{Gaussian perturbation to simulate $R$ perturbed data sets from the posterior predictive distribution.}
    \label{alg:perturb_data}
\end{algorithm}

% \begin{figure}[t!]
%     \fig{perturb_data_format2.png}{1\columnwidth}{}
%     \caption{The perturbed data sets simulated by a posterior predictive distribution of a measurement error model. The information about the measurement error is used only for simulation, and does not appear in the perturbed data sets. Thus, any traditional classification method can be trained on each perturbed data set independently, and the variation of the multiple fits can be ascribed to the measurement error uncertainty.\label{fig:data_after}}
% \end{figure}

\begin{figure}[t!]
    \centering
    \includegraphics[scale = 0.165]{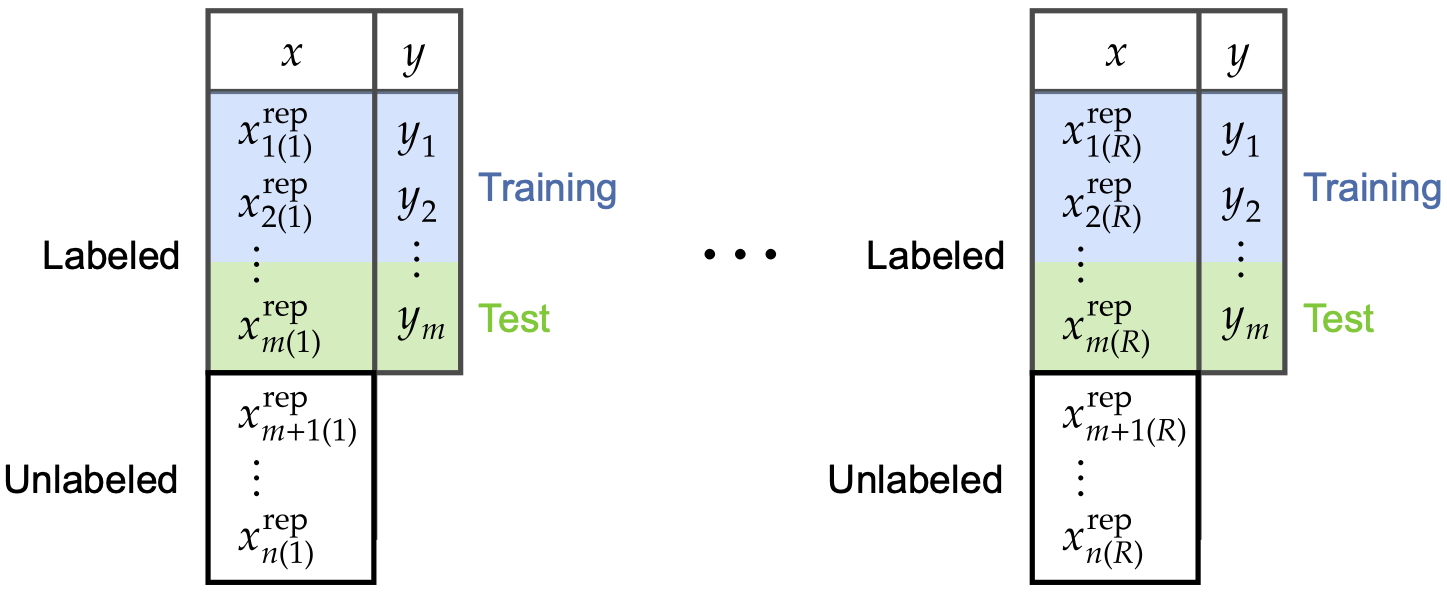}
    \caption{The perturbed data sets simulated by a posterior predictive distribution of a measurement error model. The information about the measurement error is used only for simulation, and does not appear in the perturbed data sets. Thus, any traditional classification method can be trained on each perturbed data set independently, and the variation of the multiple fits can be ascribed to the measurement error uncertainty.}    \label{fig:data_after}
\end{figure}

We point out that under the proposed framework, the information about the measurement error uncertainty is used only to generate perturbed datasets. Since the error information is incorporated into the ensemble of perturbed datasets, each individual perturbed dataset itself does not contain a column for measurement error, as shown in Figure~\ref{fig:data_after}. Therefore, any traditional classification method can be fit to each perturbed dataset as if the information about the measurement error were not given. The key point is that the variation across the multiple fits will naturally reflect the known measurement error uncertainty because the variation is \textcolor{black}{the result of} the perturbations under the Gaussian measurement error model.

\subsection{Quantifying Overall Uncertainty of a Classifier} \label{sec:overall_uncertainty}

To utilize the uncertainty propagated from measurement error, we fit a classifier to each of the $R$ simulated data sets and obtain $R$ classification results. We can then summarize the variation across the results of any quantity of interest related to the classification, such as classification accuracy and decision rule parameters.

To illustrate this, let $\theta$ denote some measure of classification accuracy, and $\theta_{(r)}$ the measured classification accuracy obtained from the labeled test set of the $r$-th perturbed data set. Then, the distribution of $\theta_{(1)}, \theta_{(2)}, \ldots, \theta_{(R)}$ obtained from $R$ independent fits represents the posterior predictive distribution of the classification accuracy $\theta$. The uncertainty of the classification accuracy $\theta$ can be quantified by the spread of this posterior predictive distribution, e.g., using the posterior standard deviation or a credible interval. Algorithm~\ref{alg:summary} provides a general template for obtaining the posterior predictive distribution for any metric $\theta$ that is a function of the classification results.  Lastly, we note that this fitting procedure is easily parallelized to reduce computational burden as each fit  on a  simulation can be performed independently.

\begin{algorithm}[t!]
    \SetAlgoLined
    \SetKwInOut{Input}{input}
    \SetKwInOut{Output}{result}

    \Input{perturbed data sets $X_{(1)}^{\mathrm{rep}}, \ldots, X_{(R)}^{\mathrm{rep}}$, classification algorithm $C$, metric~$\theta$}
    \Output{A sample, $\{\theta_{(1)}, \dots, \theta_{(R)}\}$, of size $R$ from the posterior predictive distribution of~$\theta$}
    \For{$r=1 \dots R$}{
        Fit classifier $C$ to $X_{(r)}^{\textrm{rep}}$ \\
        Calculate metric of interest $\theta_{(r)}$
     }
    \caption{The posterior predictive distribution of any classification metric~$\theta$.}
    \label{alg:summary}
\end{algorithm}

%%%%%%%%%%%%%%%%%%

\subsection{Prediction on the Unlabeled Set} \label{sec:invidivual_obs}

% hard classifiers
The proposed framework also provides a \textcolor{black}{natural way} to predict each individual's unknown label while accounting for measurement error uncertainty. For hard classifiers, we obtain a set of predicted class labels for each object from the multiple fits. Using these predicted labels, we can estimate class probabilities similar to soft classifiers. We refer to this as \textit{softening hard classifiers}. The distinction between the softened hard classifier  and traditional soft classifiers is that the former incorporates measurement error uncertainty into the estimated class probabilities. For soft classifiers, we obtain a set of estimated class probabilities for each object from multiple fits, providing additional information about the uncertainty in classifying an object. We refer to this as \textit{softening soft classifiers} (see Figure~\ref{fig:softenClassifier}).

We first discuss how to make a prediction via softening a hard classifier to obtain a single predicted label for each object in the unlabeled data set. This procedure is essentially the same as that of a traditional soft classifier. We define $K$ as the number of classes we wish to classify objects into, and 
\begin{equation*}
\hat y_{i(r)} \in \{1, 2, \dots, K\}
\end{equation*}
as the predicted class label of the $i$-th object obtained from a fit on the $r$-th perturbed data set. By conducting $R$ simulations, we obtain $R$ predictions for the $i$-th object, $\hat y_{i(1)}, \ldots, \hat y_{i(R)}$. Using these $R$ predictions, we compute the proportion of simulations that classify object $i$ into class $k$,

\begin{equation} \label{eq:obs_prob}
    %\bar y_j = \frac 1S \sum_{s = 1}^S \hat y_j^{(s)},
    \hat p_{ik} = \frac1R \sum_{r = 1}^R \mathds{1} \{\hat y_{i(r)} = k \},
\end{equation}
where $\mathds{1} \{\hat y_{i(r)} = k \}$ is an indicator equaling 1 if the $r$-th simulation places object $i$ into class $k$ and 0 otherwise. This proportion is the estimated probability that object~$i$ belongs to class $k$. 

% \begin{figure*}[t!]
%   \fig{toy_data.png}{1\textwidth}{}
%   \caption{Left: the true unobserved data. Center: the observed data without considering measurement error, suitable for standard classification methods. Right: the observed data with error bars whose lengths represent 2 standard deviations of measurement error uncertainties (i.e., measurement $\pm 1\sigma$), which the proposed framework utilizes to simulate perturbed data sets.\label{fig:truedata}}
% \end{figure*}

\begin{figure*}[t!]
  \includegraphics[scale = 0.09]{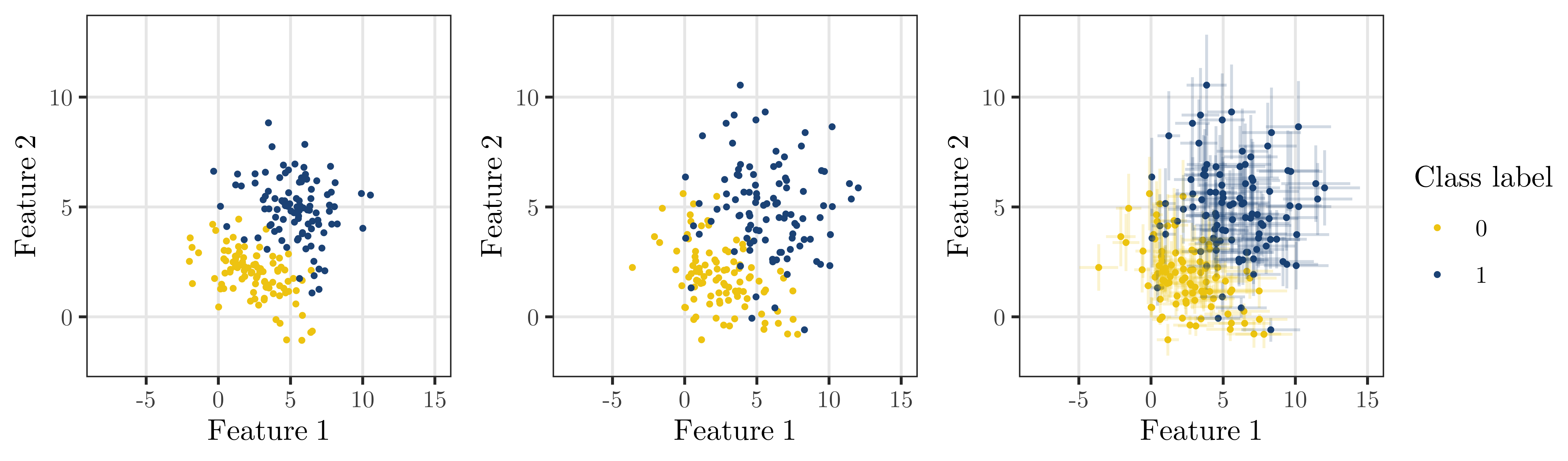}
  \centering
  \caption{Left: the true unobserved data. Center: the observed data without considering measurement error, suitable for standard classification methods. Right: the observed data with error bars whose lengths represent 2 standard deviations of measurement error uncertainties (i.e., measurement $\pm 1\sigma$), which the proposed framework utilizes to simulate perturbed data sets.}
  \label{fig:truedata}
\end{figure*}

Finally, the predicted class of the $i$-th object is $k$ if its estimated probability of being in class $k$, $\hat p_{ik}$, is greater than or equal to some decision threshold $t$ between 0 and 1. The threshold is typically chosen by scientists for their purposes \citep{he2009learning}. For example, a simple  choice for $t$ in a binary classification problem is to set $t=0.5$, which is equivalent to predicting the class of the $i$-th object as class~1 when $\hat p_{i1} > \hat p_{i0}$. One disadvantage of this simple choice is that it classifies objects into class~1 even when the classification is uncertain, e.g., with $\hat p_{i1}=0.501$ and $\hat p_{i0}=0.499$.   Thus, it is desirable to set the threshold according to the purpose of classification in practice. For example, if object~$i$ did not have estimated probabilities greater than $t=0.8$, then we might consider the class of object~$i$ as a new class, `ambiguous' (e.g., unsafe cases in \citealt{napierala2015abstaining}). This ambiguous class ensures that uncertain objects are separated from more certain ones, and that the remaining classes contain only objects that we can be confident about. Setting a higher threshold, such as $t=0.9$, will give even smaller sample in each  class but with greater purity. Such a strong decision rule may provide an  effective way to remove objects with large measurement errors from further scientific consideration. This threshold approach can also be used for a  multi-class problem with more than two classes. That is, the predicted class of object $i$ is $k$ if $\hat p_{ik}$ is the only probability estimate greater than $t$ and `ambiguous' otherwise.

For inherently soft classifiers, the output of each fit is a set of estimated class probabilities such as $\hat p_{i1}, \hat p_{i2}, \ldots, \hat p_{iK}$ for the $i$-th object instead of a single class label. An implementation of random forest, for example, will plant many  classification trees (forming a forest), make a label prediction using each tree, and compute $\hat p_{ik}$ as formulated in Equation~\eqref{eq:obs_prob}. Gaussian perturbation provides possible variations of $\hat p_{ik}$ via $R$ simulations, i.e.,  $\hat p_{ik(1)}, \hat p_{ik(2)}, \ldots, \hat p_{ik(R)}$, whose variation reflects the measurement error uncertainty. The average of these $R$ variants provides a more informative estimated class probability by accounting for measurement error, i.e.,
\begin{equation}\label{eq:phatik}
\hat{p}^+_{ik}=\frac{1}{R}\sum_{r=1}^R\hat p_{ik(r)}.
\end{equation}
We use the superscript `$+$' to  indicate that the quantity is obtained after accounting for measurement error uncertainties. We have selected this symbol because measurement error uncertainties are typically visualized by cross bars around each 2-dimensional data point, e.g., as shown in the third panel in  Figure~\ref{fig:truedata}. This formulation satisfies $\sum_{k=1}^K\hat{p}^+_{ik}=1$, ensuring that the estimated probabilities for each object sum to 1 across all classes. Finally, the  class of the $i$-th object is predicted to be $k$ if  $\hat{p}^+_{ik}$ is greater than or equal to some threshold $t$, as is usually done for soft classifiers.

\subsection{Computation and software}

Computations for this study were made with the R statistical programming language \citep{R}. Random forest and support-vector machine classification were performed with CRAN packages \textit{caret} \citep{caret} and \textit{e1071}.  Several other CRAN packages (\textit{dplyr, data.table, magrittr, tictoc, foreach, logger, ggplot2, ggpubr}) were used for data manipulation, parallelization, and graphics.  The \texttt{R} scripts are available on GitHub for reproducibility\footnote{\url{https://github.com/sarahshy/GaussianPerturbation}}.

%%%%%%%%%%%%%%%%%%%%%%%%%%%%%%%%%%%%%%%%%%%%%%%%%%%%%%%%%%%%%%
\section{Simulation Study}\label{sec:simulation}
%%%%%%%%%%%%%%%%%%%%%%%%%%%%%%%%%%%%%%%%%%%%%%%%%%%%%%%%%%%%%%

We illustrate how Gaussian perturbation can be used to better quantify classification uncertainty and make predictions accounting for measurement error uncertainty. Let us consider the following simulation setting with two features ($p=2$) and heteroscedastic measurement error uncertainties. We instantiate a set of `true' data with 200 observations by sampling from two bivariate Gaussian distributions, 
\begin{equation*}
    \begin{aligned}
        \xtrue{1}, \dots, \xtrue{100} &\sim \mathcal N \left(         
            \begin{bmatrix}
                5 \\ 5
            \end{bmatrix},
            \begin{bmatrix}
                3 & -0.5 \\ -0.5 & 2
            \end{bmatrix} \right) \\
        \xtrue{101}, \dots, \xtrue{200} &\sim \mathcal N \left( 
            \begin{bmatrix}
                2 \\ 2
            \end{bmatrix},
            \begin{bmatrix}
                4 & -1 \\ -1 & 1
            \end{bmatrix} \right) \\
    \end{aligned}
\end{equation*}
where $x_{i}^{\textrm{true}} = (x_{i1}^{\textrm{true}},~  x_{i2}^{\textrm{true}})$ for all $i$. The true class labels, denoted as $y_i$, are known from the setup,
\begin{equation*}
    \begin{aligned}
        y_i &= 0, ~~\textrm{for}~~ i=1, \dots, 100 \\
        y_i &= 1, ~~\textrm{for}~~ i=101, \dots, 200  \\
    \end{aligned}
\end{equation*}
These data represent \emph{true} values of 200 objects belonging to two classes labeled 0 and 1, and are displayed in the left panel of Figure~\ref{fig:truedata}.

To simulate noisy observations of the true data, we arbitrarily define the measurement error variance as $\sigma^2_{ij}=\vert \xijtrue\vert / 2$. These errors are heteroscedastic, where larger values of \textcolor{black}{
$\xijtrue$} are subject to greater measurement error. This type of behavior occurs, for example, in astronomical surveys where the  measured features are magnitudes. Next, we add measurement errors $\eps_{ij}$'s to the true values to obtain the \emph{observed} data $\xijobs$,
\begin{equation*}
\xijobs = \xijtrue + \eps_{ij},
\end{equation*}
where $\eps_{ij}$ is sampled from $N(0,~ \sigma^2_{ij})$. The simulated observations, $\xiobs$, are shown in the center panel of Figure~\ref{fig:truedata}, representing the typical data used in standard classification methods (i.e., without accounting for measurement error). In the right panel of Figure~\ref{fig:truedata}, the $1\sigma$ measurement error uncertainties are superimposed on the observed values, representing the information utilized by the proposed framework. These 200 observations serve as the labeled set.

\begin{figure}[t]
    \includegraphics[scale = 0.09]{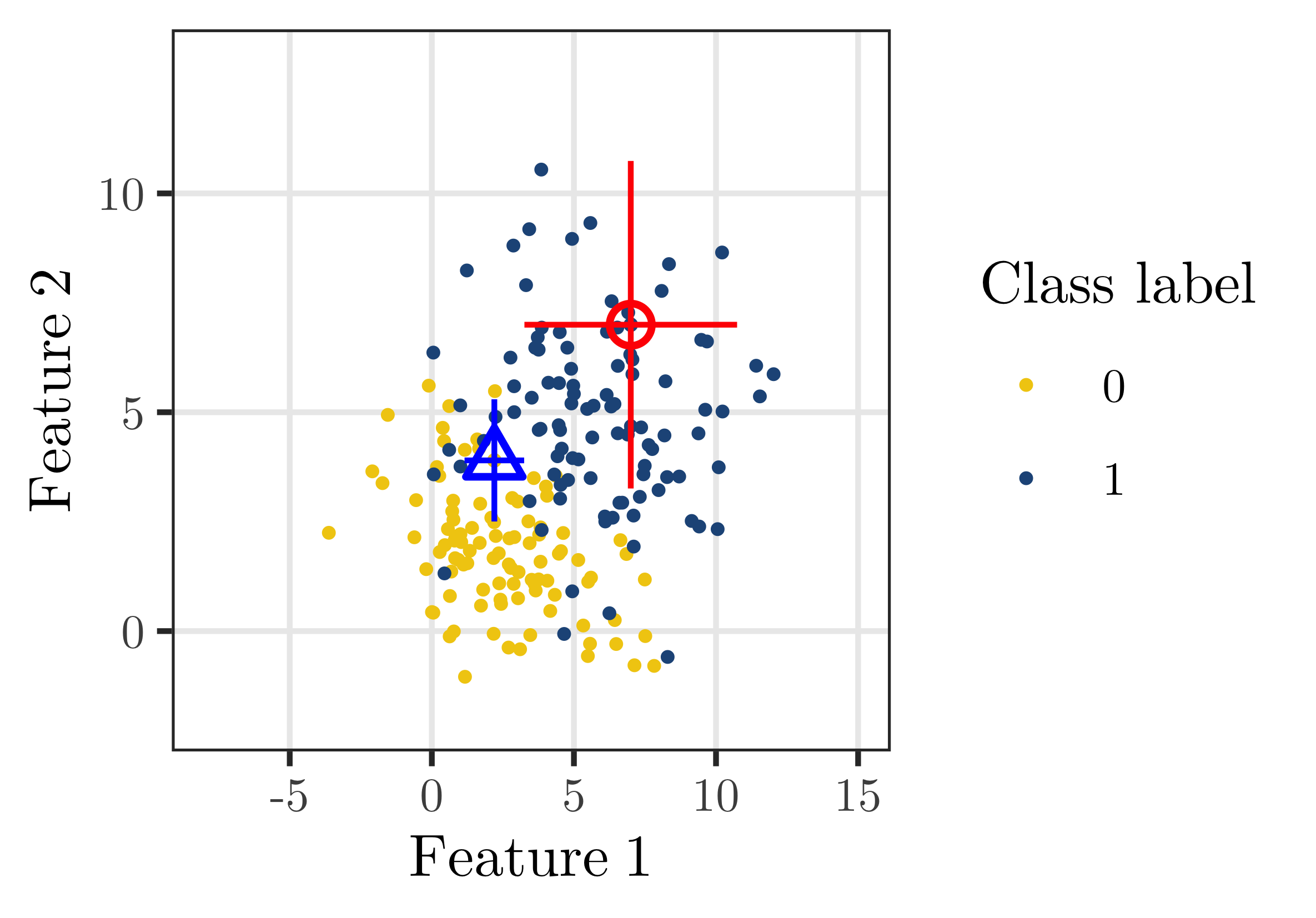} % formerly, unlabeled_points.png
    \centering
    \caption{Two observations in the unlabeled set are superimposed on the 200 observations in the labeled set.}
    \label{fig:unlabeled}
\end{figure}

In addition to the labeled set, we carefully construct an unlabeled set of two observations to illustrate how measurement error affects their class prediction  under the proposed framework. These observations are  highlighted in Figure~\ref{fig:unlabeled}; unlabeled object~1 (red circle)  lies far from the overlapping area of the two classes with a large error bar and unlabeled object~2 (blue triangle) lies near the overlapping area with a large error bar.

In this simulation study, we use a simple decision threshold $t=0.5$ because we do not have a specific scientific motivation to choose a higher threshold. That is, the predicted class of the $i$-th object is 1 if the probability of this object being in class~1 is higher than that of being in class~0 ($\hat p_{i1}>\hat p_{i0}$ or $\hat{p}^+_{i1}>\hat{p}^+_{i0}$).

\subsection{Support-vector machine}\label{sec:SVM}

For comparison, we implement a linear SVM with and without considering measurement error. The linear SVM is originally designed to utilize only the measurements $X^{\textrm{obs}}=\{x^{\textrm{obs}}_{1}, \ldots, x^{\textrm{obs}}_{202}\}$  without incorporating measurement error $\sigma=\{\sigma_{1}, \ldots, \sigma_{202}\}$. To fit this model, we use 10-fold cross validation on the labeled set (i.e., on the first 200 observations) to tune the hyper\textcolor{black}{-}parameters of the model and estimate classification accuracy. The fitted model is then used to predict the class labels of the two observations in the unlabeled set.

\begin{figure}[t!]
\includegraphics[scale = 0.105]{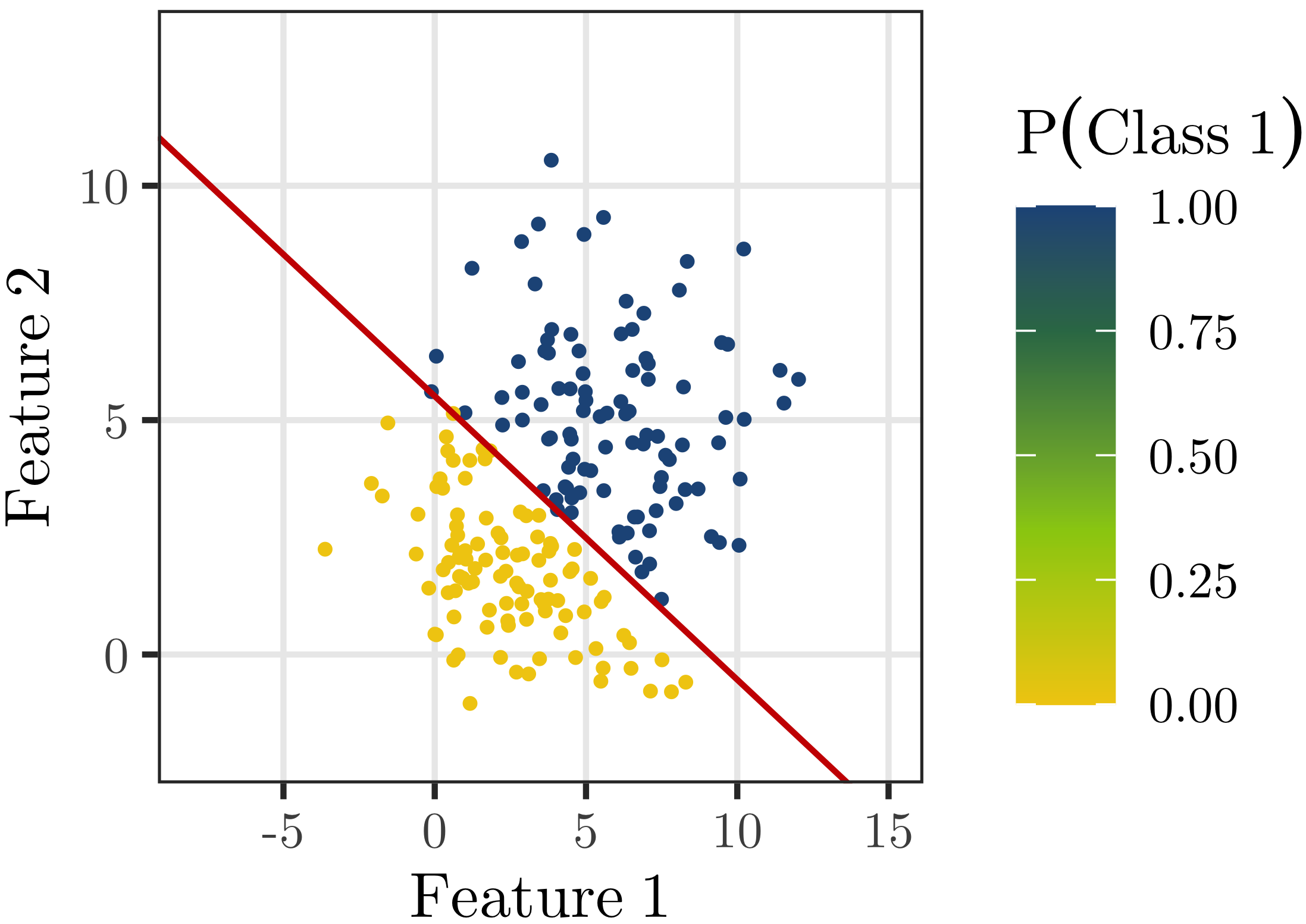}
~\\~\\
\includegraphics[scale = 0.105]{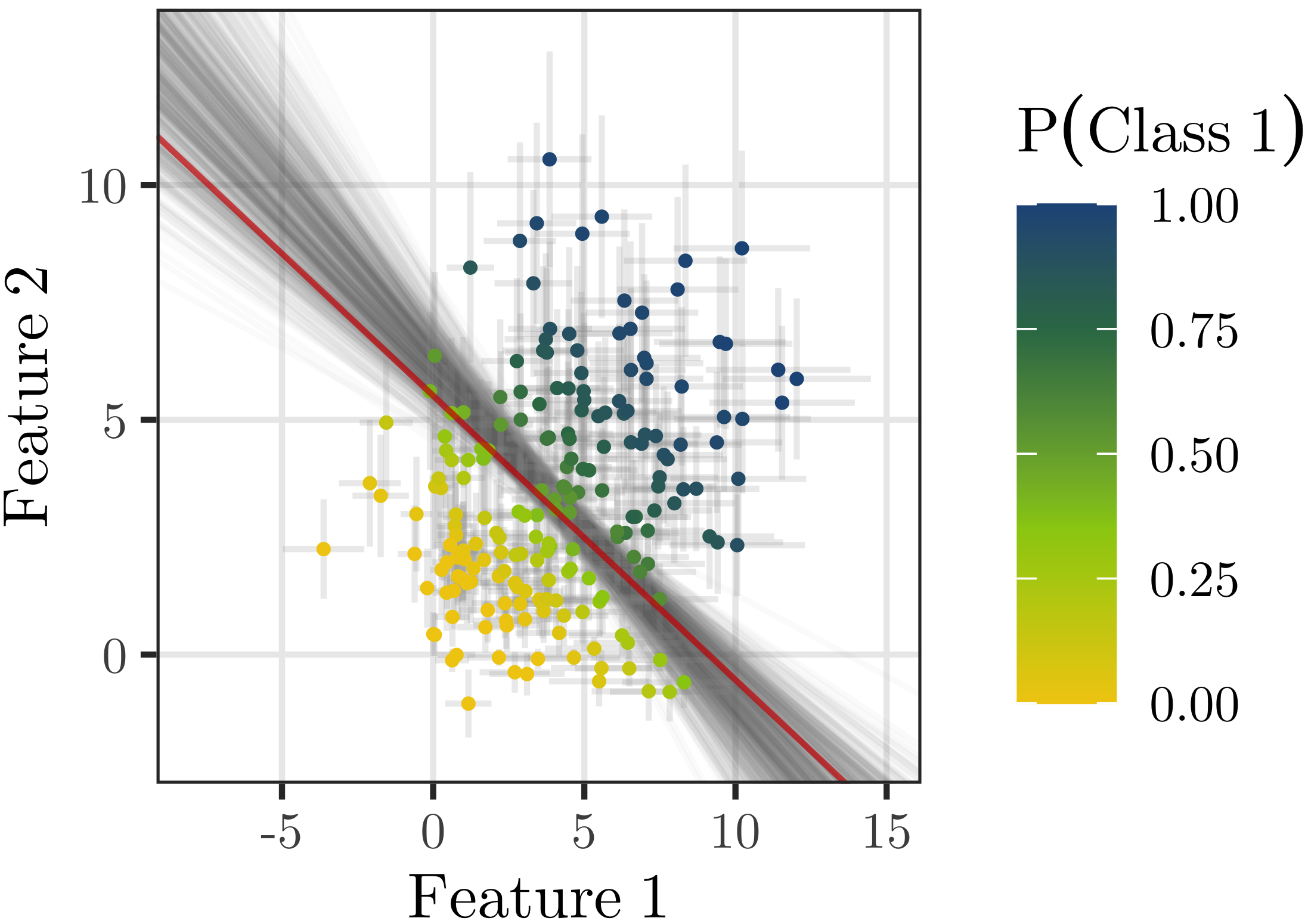}
  \centering
  \caption{Top: Linear SVM boundary fit to the observed data without accounting for measurement error, colored by the predicted labels. The probability of being in class~1 is 1 for blue dots and 0 for yellow dots. The red line is the fitted decision boundary of the linear SVM. Bottom: $500$ SVM decision boundaries obtained by fitting the SVM on each of the 500 perturbed data sets, are colored in gray. The red line is reproduced from the top panel for a comparison. The estimated probability of belonging to class 1, i.e, $\hat{p}_{i,1}$ for each object $i$ is denoted by a blue-green-yellow color gradient; darker dots are more likely to belong to class~1. Both the training and test sets are plotted, but each decision boundary was derived using the training set only. The length of each error bar is 2 standard deviations of measurement error (i.e., measurement $\pm 1\sigma$).}
  \label{fig:toysvm}
\end{figure}

We then apply Gaussian perturbation to the labeled and unlabeled sets to generate 500 simulated data sets, $X^{\textrm{rep}}_{(1)}, \ldots, X^{\textrm{rep}}_{(500)}$. For each data set, we fit the linear SVM via 10-fold cross-validation, and predict on the unlabeled set. 

The top panel of Figure~\ref{fig:toysvm} shows the labeled data and the resulting red decision boundary without accounting for measurement error. The bottom panel visualizes the result of incorporating measurement error using the proposed approach. The 500 gray decision boundaries obtained by fitting the linear SVM on each of the 500 perturbed data sets appear to form a decision `band'. This band represents the uncertainty of the decision boundary due to measurement error, i.e., the variation around the red line.

\begin{deluxetable*}{lccc}[ht]
    \tablewidth{0pt}
    \tablecolumns{4}
    \tablecaption{SVM Performance With and Without Measuremenet Error Uncertainties\label{table:SVM_results}}
    \tablehead{\multirow{2}{*}{\textbf{Method}} & \multirow{2}{*}{\textbf{Classification accuracy}} & \multicolumn{2}{c}{\textbf{SVM decision boundary}} \\
            & & Intercept & Slope}
    \startdata
            SVM without measurement error & $0.91$ & $5.51$ & $-0.60$ \\ \hline
            SVM with  measurement error  & $0.80  \pm 0.03$ & $6.15 \pm 0.60$ & $-0.71\pm 0.13$ \\  \hline
            Posterior predictive distribution &
                \raisebox{-1\totalheight}{\includegraphics[scale = 0.12]{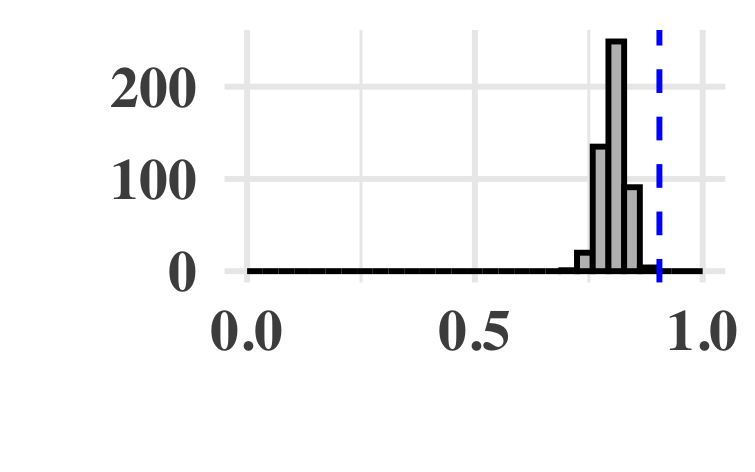}} &
                \raisebox{-1\totalheight}{\includegraphics[scale = 0.12]{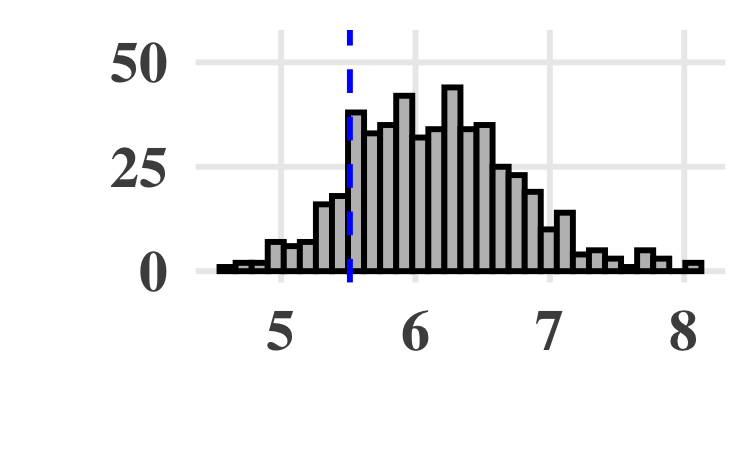}} &
                \raisebox{-1\totalheight}{\includegraphics[scale = 0.12]{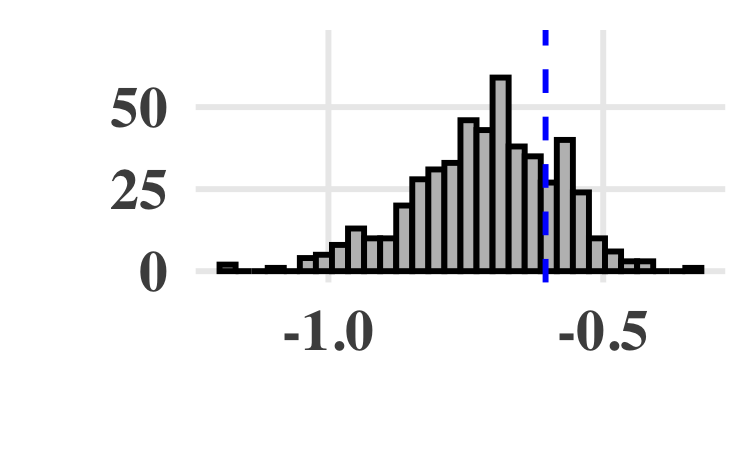}}
\enddata
\tablecomments{Comparison of overall classification accuracy and decision boundary parameters of a linear SVM computed from the labeled set with and without incorporating measurement error uncertainties. The vertical dashed lines in the posterior predictive distributions indicate the values obtained from the SVM without measurement error.}
\end{deluxetable*}

We first compare the classification accuracy between the standard approach and the proposed approach considering measurement error. Without accounting for measurement error, the cross-validated classification accuracy is 0.91. Accounting for measurement error, the average cross-validated classification accuracy is 0.80 with a standard deviation of 0.03; see Table~\ref{table:SVM_results} for details. The variation in accuracy stems from the variation in the SVM decision boundary due to measurement error. The former (0.91) is more than three standard deviations away from the accuracy obtained when accounting for measurement error (i.e., larger than 0.89). This  indicates that over\textcolor{black}{-}confidence about our measurements can lead to exaggerated results and potential bias, as is common in linear regression fits with/without measurement error \citep{akritas1996linear}. The lower accuracy of the proposed method is unsurprising as the presence of measurement error has further blurred the separation between the classes.

The estimated probability of each object belonging to class 1, $\hat{p}_{i1}$, is calculated as defined in Equation~\eqref{eq:obs_prob} and visualized using a blue-green-yellow color gradient in the bottom panel of Figure~\ref{fig:toysvm}; darker dots are more likely to belong to class 1. Objects that are closer to the decision band are less certain, with estimated class probabilities near 0.5. In the top panel of Figure~\ref{fig:toysvm}, colors are either blue or yellow without gradation because SVM predicts each object's class as a single label with probability 1.

% \begin{table*}[t!]  
%     \caption{Comparison of overall classification accuracy and decision boundary parameters of a linear SVM computed from the labeled set with and without incorporating measurement error uncertainties. The vertical dashed lines in the posterior predictive distributions indicate the values obtained from the SVM without measurement error.} 
%     \label{table:SVM_results}
%     \renewcommand{\arraystretch}{1.2}
%     \begin{tabular}{ >{\centering\arraybackslash}p{5cm}ccc }
%         \toprule
%         \multirow{2}{*}{\textbf{Method}} & \multirow{2}{*}{\textbf{Classification accuracy}} & \multicolumn{2}{c}{\textbf{SVM decision boundary}} \\
%         & & Intercept & Slope \\
%         \hline
%         SVM without measurement error & $0.91$ & $5.51$ & $-0.60$ \\ \hline
%         SVM with  measurement error  & $0.80  ~(0.03)$ & $6.15~(0.60)$ & $-0.71~(0.13)$ \\ \hline
%         Posterior predictive distribution &
%             \raisebox{-0.75\totalheight}{\includegraphics[scale = 0.12]{figures/perturbed/pert_svm_acc.png}} &
%             \raisebox{-0.75\totalheight}{\includegraphics[scale = 0.12]{figures/perturbed/pert_svm_intercept.png}} &
%             \raisebox{-0.75\totalheight}{\includegraphics[scale = 0.12]{figures/perturbed/pert_svm_slope.png}} \\
%         \bottomrule
%     \end{tabular} 
% \end{table*}

Finally, we compare the prediction results of the two observations in the unlabeled set. As shown in Figure~\ref{fig:unlabeled}, unlabeled object 1 (red circle) lies far from the intersection of the two classes and is near the majority of class~1 objects  with large measurement error. The SVM without measurement error classifies this into class~1 with probability~1. On the other hand, SVM with measurement error makes a  less certain prediction with an estimated probability of being class~1 equal to 0.78. Even for such a seemingly clean-cut object that is far from the decision band, 22\% of the  perturbed values of this object would lie below the decision boundary if its large measurement error were considered.  Object~2 (blue triangle) lies on the intersection with large measurement error. Without considering measurement error, this object's class is predicted to be class~0 with probability~1. With measurement error, however, the uncertainty of the object is reflected, with an estimated probability of belonging to class~0 equal to 0.46.

Consequently, even though single predicted labels  are consistent under both approaches, we note that the proposed approach  provides more elaborate uncertainty quantification.

%%%%%%%%%%%

\subsection{Random Forest} \label{sec:rf}

We fit a random forest on the same data of 202 objects, repeating the same 10-fold cross validation procedure. The cross-validated classification accuracy of random forest without accounting for measurement error is 0.87. Considering measurement error, the average classification accuracy decreases to 0.77 with standard deviation 0.03. This result is similar to that obtained by SVM in Section~\ref{sec:SVM}. It indicates that random forest is also sensitive to the measurement error and that ignoring measurement error can exaggerate classification performance.

% \begin{figure}[t!]
%     \includegraphics[scale = 0.105]{figures/rf1.png}
%     ~\\~\\
%         \includegraphics[scale = 0.105]{figures/rf2.png}
    
%     \centering
%     \caption{The probability estimate  of being in Class~1  computed from a one-time implementation of random forest without measurement error ($\hat{p}_{i1}$ for each object $i$, defined in Equation~\eqref{eq:obs_prob}) is visualized via color gradation in the top panel. The probability estimate calculated by the proposed approach that accounts for measurement error ($\hat{p}^+_{i1}$ for each object $i$, defined in Equation~\eqref{eq:phatik}) is visualized via the same color gradation in the bottom panel. Darker dots are more likely to belong to class 1.}
%     \label{fig:rfprob}
% \end{figure}

\begin{figure}[t!]
    \gridline{\fig{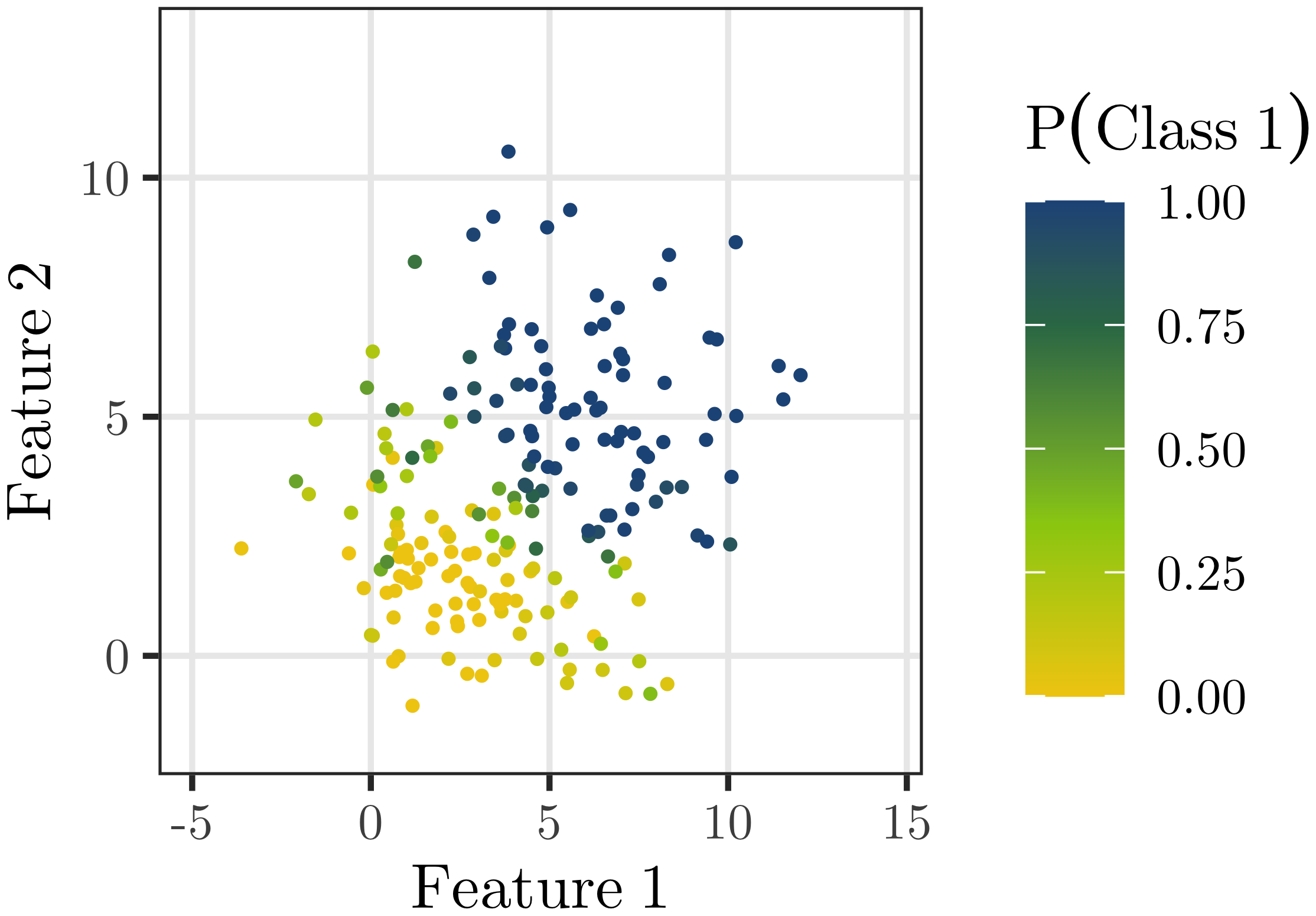}{0.95\columnwidth}{}}
    \gridline{\fig{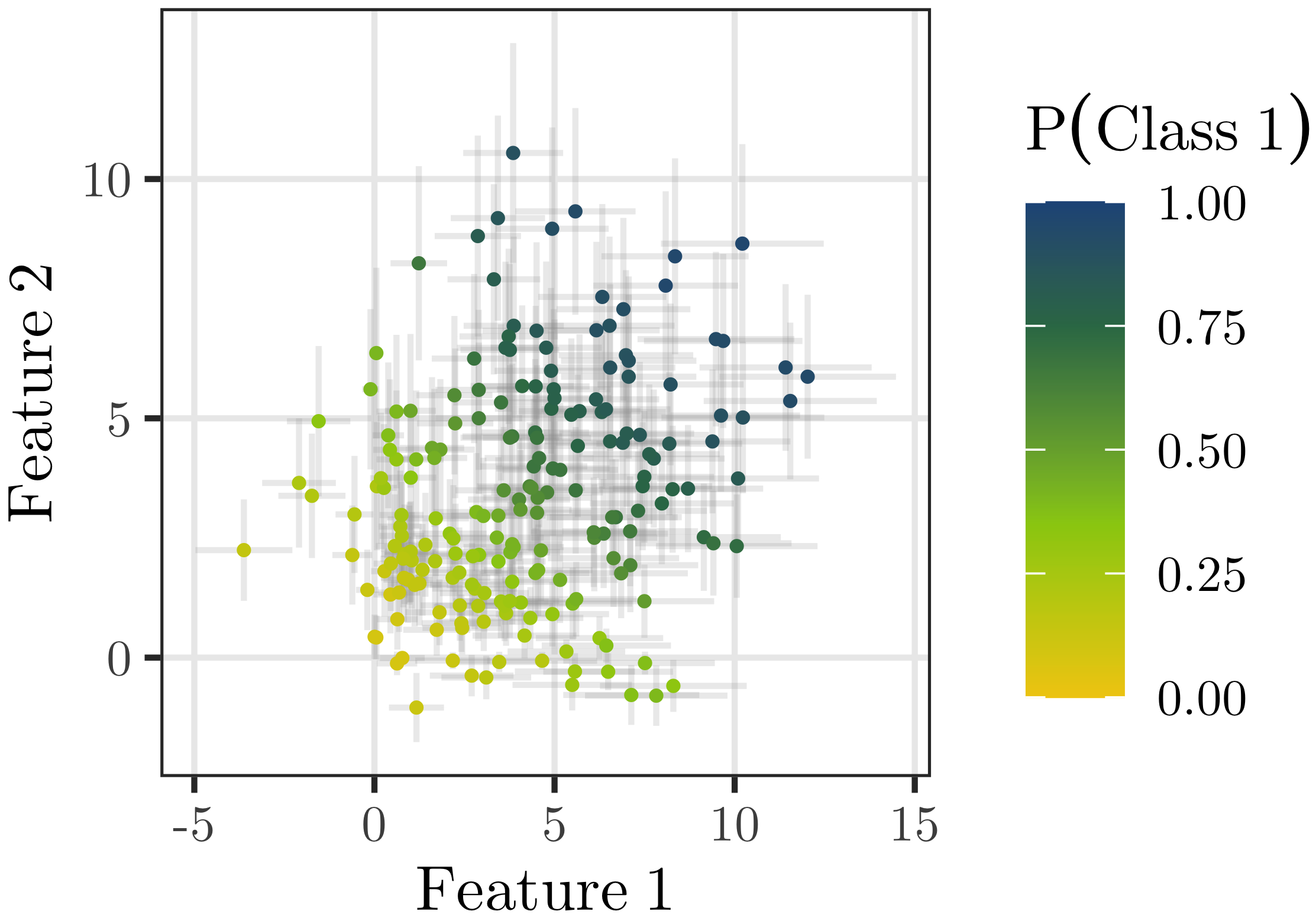}{0.95\columnwidth}{}}
    \caption{The probability estimate of being in Class~1  computed from a one-time implementation of random forest without measurement error ($\hat{p}_{i1}$ for each object $i$, defined in Equation~\eqref{eq:obs_prob}) is visualized via color gradation in the top panel. The probability estimate calculated by the proposed approach that accounts for measurement error ($\hat{p}^+_{i1}$ for each object $i$, defined in Equation~\eqref{eq:phatik}) is visualized via the same color gradation in the bottom panel. Darker dots are more likely to belong to class 1.\label{fig:rfprob}}
\end{figure}

Next, we visually compare the estimated probabilities obtained with and without measurement error. The top panel of Figure~\ref{fig:rfprob} displays the estimated probabilities of belonging to class~1 obtained without measurement error ($\hat{p}_{i1}$ for each object $i$, defined in Equation~\eqref{eq:obs_prob}) using the same color gradient as in Figure~\ref{fig:toysvm}. Random forest clearly shows some color gradation near the overlapping area, unlike SVM's strict blue-yellow distinction around the decision boundary. However, the gradient is sporadic rather than continuously changing near the overlapping area, with dark green dots scattered even in the predominantly yellow area of class 0 (i.e., near the lower-left corner).  It turns out that these dark green dots in the yellow-dominant region are known to belong to class~1. A comparison between the left and middle panels of Figure~\ref{fig:truedata} indicates that these data \textcolor{black}{have} become isolated from class~1 while simulating noisy measurements.  From this result, one may conclude that random forest without measurement error has worked very well, identifying these dark green dots even deep inside the yellowish area.  However, we point out that this standard random forest approach is not concerned with measurement error, i.e., why these data became isolated from the majority of class~1 in the beginning. Instead, it assumes that the data are perfectly measured. It has just worked well under this assumption, producing relatively high probabilities of these data belonging to class~1, even though their \emph{error-free} measurements put them near the majority of class~0.

On the other hand, the proposed approach reflects the possibility that these class~1 data may be located near the majority of class~0 simply due to measurement error. The estimated probabilities obtained by Gaussian perturbation with measurement error, i.e., $\hat{p}^+_{i1}$  in Equation~\eqref{eq:phatik}, are visualized in the bottom panel of  Figure~\ref{fig:rfprob} using the same color gradient. Compared to the top panel (obtained without measurement error), the bottom panel shows a thick greenish band. This band occupies most of the previously yellow-dominant region, leaving a smaller yellow area in the lower-left corner. This indicates that classifying objects into class~0 has become less certain after we account for measurement error. This phenomenon can be ascribed to the isolated class~1 points residing within the majority of the class~0 data. Since each of these isolated class~1 data is used for training in 9 folds during 10-fold cross-validation, the trees of a random forest routinely identify nearby data to be class~1. When aggregated, this increases the probability of being in class~1 for the nearby class~0 data. The act of perturbing the data using Gaussian perturbation spreads this uncertainty deeper into the lower-left corner. For example, half of perturbed values of the isolated class~1 data are generated even further away from class~1 (i.e., closer to the lower-left corner), and each of them increases the probability of nearby data being in class~1. Thus, due to the average effect of the proposed approach, the previously yellow-dominant area has become greenish, reflecting more uncertainty in classifying objects into class~0.

Lastly, we predict the classes of the two observations in the unlabeled set. Unlabeled object 1 (red circle in Figure~\ref{fig:unlabeled}) is  near the majority of class 1 objects  with large measurement error. Random forest without measurement error classifies this into class~1 with an estimated probability of being in class~1 ($\hat{p}_{i0}$) equal to~1. With measurement error incorporated, the  probability estimate ($\hat{p}^+_{i1}$) decreases to 0.79 with standard deviation 0.29, appropriately reflecting its large measurement error. It turns out that both approaches predict this unlabeled object to be in class~1 with a simple  threshold of $t=0.5$ (i.e., $\hat{p}_{i1}>\hat{p}_{i0}$ and $\hat{p}^+_{i1}>\hat{p}^+_{i0}$ for $i=201, 202$). But we note that the proposed approach provides the more comprehensive information about its prediction uncertainty.

Object 2 (blue triangle  in Figure~\ref{fig:unlabeled}) is located in the overlapping area with large measurement error. Without considering measurement error, this object's class is predicted to be class~1 with estimated probability $\hat{p}_{i1}=0.51$.  This probability implies that our class prediction for this observation is as good as a coin toss. By considering measurement error, however, the proposed approach reveals ambiguity in a more elaborate way, producing the probability estimate $\hat{p}^+_{i1}=0.45$ with standard deviation 0.32. Consequently, the label prediction  for this  object becomes completely different, i.e., predicted to be class~1 without measurement error and class~0 with measurement error. It implies that measurement error can play a significant role in predicting unknown labels in practice. We also note that while the estimated probabilities are similar, the proposed approach tells more about the uncertainty involved in the prediction  than a simple coin toss.

%%%%%%%%%%%%%%%%%%%%%%%%%%%%%%%%%%%%%%%%%%%%%%%%%%%%%%%%%%%%%%
\section{Classifying High-Redshift  Quasars} \label{sec:astro}
%%%%%%%%%%%%%%%%%%%%%%%%%%%%%%%%%%%%%%%%%%%%%%%%%%%%%%%%%%%%%%

Following the work of \cite{timlin2018clustering}, we consider the problem of identifying high-$z$ quasar candidates ($2.9  \leq z \leq 5.1$) in a catalog data set merged from multiple sources. Here we adopt only the random forest classifier to demonstrate the application of the Gaussian perturbation framework, although other classifiers can be similarly applied to the dataset. Also, we adopt the same decision threshold as \citeauthor{timlin2018clustering}, setting the threshold $t=0.5$ as is done in our simulation study. 

\subsection{The Labeled and Unlabeled Data Sets}
We collect labeled and unlabeled photometric catalog data from several sources, following \citet{timlin2018clustering}.   Their scientific goal is to identify high-$z$ quasar candidates from the unlabeled set when the prevalence rate of high-$z$ quasars in the labeled set is only  ${\sim}3\%$.  The labeled data consist of optical photometric data from the Sloan Digital Sky Survey (SDSS) combined with data from the Spitzer IRAC Equatorial Survey \citep[SpIES;][]{ApJS2016spies} and the Spitzer-HETDEX Exploratory Large-Area survey \citep[SHELA;][]{ApJS2016shela}. The combined data are restricted to Stripe 82 objects \citep{annis2014sloan, jiang2014sloan} for which deep optical photometry is available in the five optical SDSS filters \citep[\emph{ugriz};][]{fukugita1996sloan}. \textcolor{black}{All magnitudes have been
appropriately corrected for Galactic extinction.}  Photometric errors for each object in each band are also provided.  \textcolor{black}{Using these raw data, we define six quasar colors (or magnitude differences),  that is, $ug, gr, ri, iz, zs1$, and $s1s2$. These six colors are the features to be used for a classification. We appropriately convert the measurement error uncertainties in magnitudes to those in colors using a Delta method.}

The 5,487 quasars in the labeled set are spectroscopically confirmed as high-$z$ quasars from the quasar catalog of \cite{richards2015bayesian}. The remaining 643,952 objects in the labeled set, consisting of stars, galaxies, and unclassified quasars, are collapsed into a single class called `AE' (Anything Else). The unlabeled set is assembled using matched optical+MIR photometric data restricted to Stripe 82 and consists of 1,862,968 objects.  See \citeauthor{timlin2018clustering}~for further details regarding how the labeled and unlabeled sets are assembled. We note that  the labeled and unlabeled data sets of this work are not exactly the same as those in \citeauthor{timlin2018clustering}~due to different time of data collection. 

% Spies: \url{https://ui.adsabs.harvard.edu/abs/2016ApJS..225....1T/exportcitation}
% Shela: \url{https://ui.adsabs.harvard.edu/abs/2016ApJS..224...28P/exportcitation}

\subsection{Classification Accuracy on the Labeled Set}

Using these labeled and unlabeled observations with their measurement error uncertainties, we first generate $R=500$ simulated data sets via Gaussian perturbation.  In each simulation,  we  apply  10-fold cross validation to the labeled data set to train and test a random forest classifier. Considering the class imbalance (only about 3\% of objects are high-$z$ quasars), we  use stratified sampling to form 10  cross validation sets, each with 3\% high-$z$ quasars and 97\% AE. That is, we randomly divide the labeled set of high-$z$ quasars and that of AE into 10 equally-sized pieces, and pair one piece of high-$z$ quasar with one piece of AE.

To assess the classification accuracy in the presence of substantial class imbalance in the data, we use two measures, \textit{completeness} and \textit{efficiency} \citep{timlin2018clustering}. This is because the typical measure for classification accuracy (the total number of correct classifications divided by the total number of  objects) is dominated by the result of the much larger AE group. Completeness --- also commonly known as sensitivity, recall, or true positive rate --- evaluates the number of correctly identified high-$z$ quasars out of all known high-$z$ quasars in the test set. Efficiency --- also referred to as precision or positive predictive value --- evaluates the number of correct high-$z$ quasar classifications out of the total number of objects classified as high-$z$ quasar.

\begin{figure}[t!]
 \includegraphics[scale = 0.18]{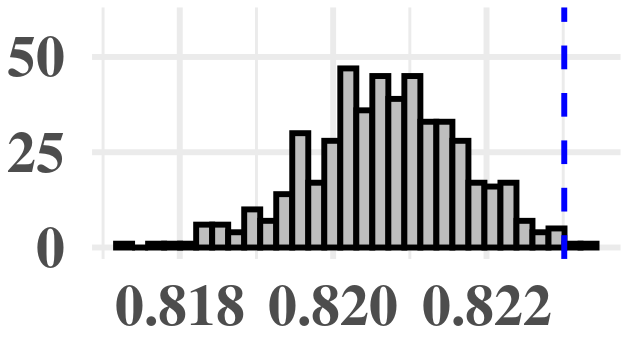}~~ \includegraphics[scale = 0.18]{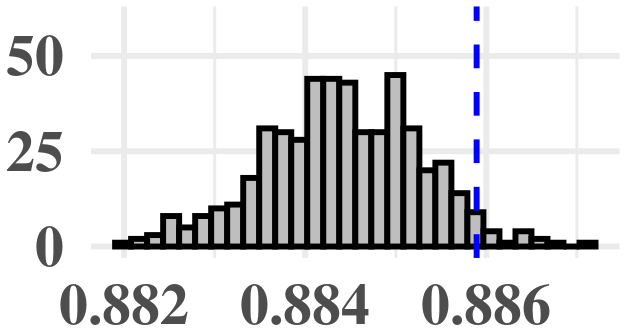}
 \centering
     \caption{Posterior predictive distributions of completeness (left) and efficiency (right), respectively, obtained by Gaussian perturbation with $500$ simulations. The mean and standard deviations are superimposed in each panel. The vertical dashed line indicates the value of completeness or that of efficiency obtained without considering measurement error.}
   \label{figure:comp_eff}
\end{figure}

In terms of these two measures, the single-run random forest without measurement error shows evidence of over-confidence in its classification accuracy. Without considering measurement error,  completeness is 82.3\% based on a single implementation of random forest. Incorporating measurement error via Gaussian perturbation, completeness  becomes 82.1\% on average across 500 simulations with standard deviation 0.1\%. The value of completeness without measurement error is on the upper bound of the two standard deviation range obtained from the proposed approach with measurement error. Thus, the outcome without measurement error can be considered nearly over-confident in completeness. Similarly, efficiency is 88.6\% without measurement error and 88.4\% $\pm$ 0.08\% with measurement error. The value of efficiency without measurement error is  higher than the two standard deviation range obtained with measurement error, indicating that the former is over-confident in efficiency. We display these results in Figure~\ref{figure:comp_eff}. The vertical dashed line in each panel  indicates the value of completeness or efficiency obtained from the one-time implementation of random forest without considering measurement error. 

%, which we believe contributes to a better astronomical science
\textcolor{black}{Without considering statistical significance, one may think that the net change in completeness or efficiency is quite modest, i.e., 0.2\%. The resulting scientific discovery based on both classification outcomes may  look similar because their net change is small (again, without considering statistical significance). We note  that the goal of Gaussian perturbation is to present more reliable and informative classification results by incorporating uncertainty from measurement error. A scientific discovery based on a classification result with  82.3\% completeness and 88.6\% efficiency is clearly less reliable than that with 82.1\% $\pm$ 0.1\% completeness and 88.4\% $\pm$ 0.08\% efficiency. This is because the latter classification outcomes mean that one makes an extra effort to account for the effect of measurement error, while the former leaves the  effect of measurement error unknown.}
%Such reliability of scientific discovery may not be obtained simply by checking `the net difference'
% However, the scientific findings based on the latter classification outcomes must be more reliable because  informative because  more complete research, investigating both possibilities with and without measurement error. 
%the former completely ignores the effect of measurement error, while the latter makes an extra effort to take into account the effect of measurement error, even if the net difference may be small. T
% though their net changes are small. This is because random forest does not have a built-in functionality to incorporate measurement error by default. This person may be happy with this classification result and  corresponding astronomical discoveries, and may be able to publish a paper  without knowing what would happen if the measurement error were considered. A better scientific paper may do more research and report both outcomes with and without measurement error as we stated in the previous paragraph (or at least the outcome based on measurement error as Bovy et al.~(2011, 2012) do), even though the net change is small (i.e., without considering statistical significance).  The resulting scientific discovery in this case may not sound very different from the one without measurement error (again, without considering statistical significance)

To discuss false-positive and false-negative rates, we display confusion matrices in Table~\ref{table:test_cm_rf} obtained with and without accounting for measurement error on the labeled set. The confusion matrix obtained via Gaussian perturbation is averaged over 500 confusion matrices, one from each simulation. The single run of random forest identifies 2,392 false-positives and 3,705 false-negatives, shown in the upper-right and lower-left cells, respectively. On the other hand, the proposed approach via Gaussian perturbation shows  2,214.9 $\pm$ 18.7 false-positives and 3701.1 $\pm$ 20.7 false-negatives when measurement error is considered.

The first column of the two tables are nearly identical; almost the same number of true high-$z$ quasars are misclassified to AE (false-negatives) regardless of whether we consider measurement error or not. This implies that these misclassified high-$z$ quasars might be located deep inside the feature space of AE objects even after accounting for the measurement error. In the second column, however, more than 100 true AE objects are misclassified as high-$z$ quasars (false-positives) without measurement error, while correctly identified as AE with measurement error. These AE objects may be distant from the majority of AE objects in the feature space, but not as distant if we consider their measurement error uncertainties. From a practical standpoint, we also note that reducing the number of false positives is extremely important  for follow-up observations. This is because a reduction of even $\sim5\%$ of false-positives is not only useful for particular scientific goals, but also saves time, cost, and effort when further investigating the candidates.

\begin{deluxetable}{crr}[t!]
\tablewidth{0pt}
\tablecolumns{3}
\tablecaption{SDSS quasar classification confusion matrices\label{table:test_cm_rf}}
\tablehead{&&} 
\startdata
\multicolumn{3}{c}{(a) Without measurement error} \\
       \textbf{Predicted\textbackslash True}  & High-$z$ Quasars  & AE \\ \hline
        High-$z$ Quasars & 16,935     & 2,392  \\
        AE               & 3,705      & 626,407  \\
&&\\
\multicolumn{3}{c}{(b) With measurement error} \\
        \textbf{Predicted\textbackslash True} & High-$z$ Quasars   & AE \\ \hline
        High-$z$ Quasars  & $16,938.9 \pm 20.7$      & $2,214.9 \pm 18.7$ \\
         AE & $3,701.1 \pm 20.7$      & $626,584.1 \pm 18.7$ \\
\enddata
\tablecomments{For case (b) the outcome is the average of 500 confusion matrices with standard deviations. The acronym `AE' denotes the class `Anything Else'.}
\end{deluxetable}

% \begin{table}[t!]
%     \caption{The confusion matrices obtained (a) without and (b) with measurement error. For case (b) the outcome is the average of 500 confusion matrices, and the values in the parentheses are standard deviations. The acronym `AE' denotes the class `Anything Else'.}
%     \label{table:test_cm_rf}
%     \begin{center}
%     (a) Without measurement error\\~\\
%     \begin{tabular}{crr}
%       \textbf{Predicted\textbackslash True}  & High-$z$ Quasars  & AE \\ \hline
%         High-$z$ Quasars & 16,935     & 2,392  \\
%         AE               & 3,705      & 626,407  \\
%     \end{tabular}~\\~\\~\\
%     (b) With measurement error\\
%     \begin{tabular}{crr}
%         \textbf{Predicted\textbackslash True} & High-$z$ Quasars   & AE \\ \hline
%         High-$z$ Quasars  & 16,938.9 (20.7)      & 2,214.9 (18.7) \\
%          AE & 3,701.1 (20.7)      & 626,584.1 (18.7) \\
%     \end{tabular}
%     \end{center}
% \end{table}

\subsection{Prediction on the Unlabeled Set via Gaussian Perturbation}

We also predict  labels of the newly observed objects in the unlabeled set with and without measurement error. To account for measurement error, we compute  estimated probabilities of being high-$z$ quasars using Gaussian perturbation, $\hat{p}^+_{i1}$ in Equation~\eqref{eq:phatik} for $i=1, 2, \ldots, n$. Then, we classify the $i$-th object in the unlabeled set  as a high-$z$ quasar if $\hat{p}^+_{i1}>0.5$ (or equivalently $\hat{p}^+_{i1}>\hat{p}^+_{i0}$).

\begin{deluxetable}{crr}[b!]
    \tablewidth{90pt}
    \tablecolumns{3}
    \tablecaption{SDSS classification for single $vs.$ ensemble random forest\label{table:prediction_catalog_rf_gp}} 
    \tablehead{&&}
    \startdata
            \textbf{Proposed \textbackslash Standard} & High-$z$ Quasars  & AE \\ \hline
                                  High-$z$ Quasars    & 8,701             & 936 \\
                                                 AE   & 3,146             & 1,850,185 \\
    \enddata
    \tablecomments{Overlap on the prediction set between a single random forest and an ensemble random forest using $R=500$ simulations. The approach without measurement error is denoted by `Standard' and the one with measurement error by `Proposed'. For example, 936 objects are predicted to be high-$z$ quasars with measurement error (Proposed), but to be AE without measurement error (Standard).  Also, 3,146 objects are classified as high-$z$ quasars without measurement  error  (Standard) but  as  AE  with  measurement  error (Proposed).}
\end{deluxetable}

% \begin{table}[b!]
%     \centering
%     \caption{Overlap of label predictions on the prediction set between a single random forest and an ensemble random forest using $R=500$ simulations. The approach without measurement error is denoted by `Standard' and the one with measurement error by `Proposed'. For example, 936 objects are predicted to be high-$z$ quasars with measurement error (Proposed), but to be AE without measurement error (Standard).  Also, 3,146 objects are classified as high-$z$ quasars without measurement  error  (Standard) but  as  AE  with  measurement  error (Proposed).}
%     \label{table:prediction_catalog_rf_gp}
%     \begin{tabular}{c rr}
%         \textbf{Proposed \textbackslash Standard} & High-$z$ Quasars  & AE \\ \hline
%                               High-$z$ Quasars    & 8,701             & 936 \\
%                                              AE   & 3,146             & 1,850,185 \\
%     \end{tabular}
% \end{table}

Table~\ref{table:prediction_catalog_rf_gp} summarizes these results. It shows that both approaches are consistent in predicting high-$z$ and AE objects on the diagonal cells. However, 936 objects are  predicted to be high-$z$ quasars with measurement error, but to be AE without measurement error. Also, 3,146 objects are classified as high-$z$ quasars without measurement error but as AE with measurement error.  Consequently, a single run of random forest without measurement error identifies $11,847$ high-$z$ quasars from $\sim$2 million objects in the unlabeled set (i.e., the sum of the first column). Taking measurement error into account,  random forest via Gaussian perturbation identifies 9,637 high-$z$ quasars (i.e., the sum of the first row), which is substantially lower than the number of high-$z$ quasars identified without measurement error.

This difference reveals important aspects of the classification with and without measurement error. The top-left cell of Table~\ref{table:prediction_catalog_rf_gp} shows that the two approaches are consistent in predicting 8,701 objects as high-$z$ quasars. However, the top-right cell indicates that 936 object are classified as AE without measurement error, while classified as high-$z$ quasars after considering measurement error. In other words, these 936 objects are potential candidates for high-$z$ quasars  \textcolor{black}{that might have been buried in a haystack of 1.85 million AE objects without considering measurement error}. Also the bottom-left cell of Table~\ref{table:prediction_catalog_rf_gp} indicates that 3,146 objects, which are classified as high-$z$ quasars without measurement error, are classified as AE after accounting for measurement error. This means that these 3,146 objects might be potential misclassifications (i.e., not high-$z$ quasars) of the standard approach that does not account for  measurement error. 

%previously hidden under the assumption that the measurement has no uncertainty

%%%%%%%%%%%%%%%%%%%%%%%%%%%%%%%%%%%%%%%%%%%%%%%%%%%%%%%%%%%%%%
\section{Discussion} \label{sec:discussion}
%%%%%%%%%%%%%%%%%%%%%%%%%%%%%%%%%%%%%%%%%%%%%%%%%%%%%%%%%%%%%%

\subsection{Why Bayesian posterior predictive distribution?}

Under the Gaussian measurement error model in Equation~\eqref{obs_dist}, the observed data are measurements of the unknown true values with Gaussian noise, i.e., 
\begin{equation*}
\xijobs \sim \mathcal{N}(\xijtrue,~ \sigma_{ij}^2),
\end{equation*}
where $\sigma_{ij}^2$ is known for all $i$ and $j$.

Based on this model, one  way to simulate replicates of the current data is to add Gaussian noise to the observation with \emph{the same} measurement error uncertainty \textcolor{black}{\citep{2008ApJ...683...12B}}. That is,
\begin{equation*}
x_{ij}^{\textrm{rep}} \sim \mathcal{N}(\xijobs,~ \sigma_{ij}^2).
\end{equation*}
\textcolor{black}{However, } under the Gaussian measurement error model, it makes more sense for replicates to be distributed around $\xijtrue$. This is because the \textcolor{black}{Gaussian measurement error} model assumes that (hypothetical) repeated measurements under the same condition are distributed as a Gaussian distribution centered at $\xijtrue$.  \textcolor{black}{Thus,} the replicates obtained  \textcolor{black}{ by this} approach would be consistent with \textcolor{black}{the Gaussian measurement error model} only if $\xijobs=\xijtrue$. In practice,  this condition is impossible to meet due to non-zero measurement error ($\sigma_{ij}>0$), violating the key idea of the Gaussian measurement error model.

A Bayesian approach provides a straightforward way to simulate replicates centered at $\xijtrue$, avoiding the inconsistency of the \textcolor{black}{previous} approach. By setting a prior distribution on $\xijtrue$, we can  derive and sample the resulting posterior distribution given the observed data $\xijobs$, i.e., $\pi(\xijtrue\mid \xijobs)$. This posterior distribution captures all possible variations of $\xijtrue$ given the data. Then, we can easily generate a replicate from a Gaussian distribution centered at each possible realization of $\xijtrue$ given the data. \textcolor{black}{This can be done by sampling $\xijtrue$ from $\pi(\xijtrue\mid \xijobs)$, and then by  sampling $x_{ij}^{\textrm{rep}}$ from $f(x_{ij}^{\textrm{rep}} \mid \xijtrue)$ given the previously sampled $\xijtrue$}. This \textcolor{black}{approach} is consistent with the assumption behind the Gaussian measurement error model. In addition, by accounting for all possible values of the unknown true value, $\xijtrue$, the uncertainty of $\xijtrue$ is naturally reflected in the replicates. This is what the following posterior predictive distribution  does by encoding the uncertainty of $\xijtrue$ into the replicates given the observed data:
\begin{equation*}
q(x_{ij}^{\textrm{rep}} \mid \xijobs)=\int f(x_{ij}^{\textrm{rep}} \mid \xijtrue)~\pi(\xijtrue\mid \xijobs)~d\xijtrue.
\end{equation*}

We note that the proposed Gaussian perturbation in Equation~\eqref{eq:ppd} is a specific posterior predictive distribution obtained with the improper flat prior distribution on $\xijtrue$. Different priors on $\xijtrue$ lead to different posterior predictive distributions  with possibly more complicated sampling steps. 

A potential extension of the current work is to adopt a proper prior distribution on $\xijtrue$ based on physical knowledge. A two-level Gaussian hierarchical model \citep{efron1975data} is an example of modeling a population distribution of each known class via a Gaussian prior distribution. For example, the unknown true values $\xijtrue$ in class~1 are assumed to be from one Gaussian population distribution, and those in class~2 from another Gaussian population distribution, etc. However, this elaborate modeling approach requires more computation prior to perturbation by sampling the more complicated posterior distribution of $\xijtrue$ before we sample the replicates $x_{ij}^{\textrm{rep}}$'s.

\subsection{\textcolor{black}{Sources of uncertainty in Gaussian perturbation}}

\textcolor{black}{The Gaussian perturbation framework has two main sources of uncertainty; measurement error uncertainty and modeling uncertainty. This is because the variation across the perturbed data sets results directly from these two types of uncertainty. The posterior predictive distribution of a replicate observation, which is used to simulated perturbed data sets, clearly shows these two sources:
$$
x_{ij}^{\textrm{rep}}\mid x_{ij}^{\textrm{obs}}\sim \mathcal{N}(x_{ij}^{\textrm{obs}},~ 2\sigma^2_{ij}).
$$
Here, the measurement error uncertainty $\sigma_{ij}$ determines the variance. Besides, we note that our Bayesian modeling assumptions are composed of the Gaussian measurement error model (likelihood) and improper flat priors on the unknown true feature values $x_{ij}^{\textrm{true}}$. These modeling assumptions make this posterior predictive distribution a Gaussian and determine the variance inflation factor~2.}

\textcolor{black}{When we compare different Bayesian classifiers, it is important to compare the modeling uncertainty. This is because  the measurement error uncertainty is completely known and is given in the data set. For example, a Bayesian classifier in \cite{2011ApJ...729..141B} also adopts the same Gaussian measurement error model. As for priors, even though they do not specify a joint prior distribution of the unknown true feature values, they mention that the resulting posterior distribution $\pi(x_{ij}^{\textrm{true}}\mid x_{ij}^{\textrm{obs}})$ can be Gaussian. The fact that both works result in a Gaussian posterior distribution of the unknown true feature value does not mean that their modeling uncertainties are the same. This is because there are possibly many priors that lead to a Gaussian posterior, and the resulting posterior variances  can differ according to the choice of prior.  Therefore, it is not possible to compare the modeling uncertainties of the two Bayesian models directly. We note, however, that our improper flat prior on each $x_{ij}^{\textrm{true}}$ is a conservative choice that reflects our lack of knowledge about these true feature quantities. This is because the resulting posterior variance $\sigma^2_{ij}$ in Equation~\eqref{eq:post_dist} is greater than any other possible Gaussian posterior variances based on an informative unimodal prior distribution.}

\subsection{The number of perturbed data sets}

One important question regarding the proposed Gaussian perturbation is  how we determine the number of perturbed data sets, $R$, in practice. In principle, the larger the value of $R$ is, the smaller the resulting Monte Carlo error  is.  This is because the Monte Carlo error converges to zero at rate $O(1/\sqrt{R})$  \citep[chap.~1.1]{liu2008monte}. Thus, it needs to be chosen large enough to capture the shape of the posterior predictive distribution without missing any important feature of the distribution, e.g., heavy-tailedness or multimodality. 

In the bootstrapping literature, it is  known that $200$ replicates of the data are typically needed and even $25$ are usually informative \citep[pp.~48 and 52]{efron1994introduction}. Since the current work is a resampling method similar to parametric bootstrapping, setting $R=200$ might be large enough. In our numerical studies, we set $R=500$ because 500 pseudo-data sets have been enough to capture the shapes of posterior predictive distributions, as shown in Table~\ref{table:SVM_results} and Figure~\ref{figure:comp_eff}.  Though not reported here, we have confirmed that more perturbed data sets (e.g., $R=1000$) do not  change the shape of each posterior predictive distribution meaningfully.

\subsection{\textcolor{black}{Classifiers  for Gaussian perturbation}}
\textcolor{black}{The proposed Gaussian perturbation is designed to incorporate the information about  measurement error  into a classifier that does not have a natural way to reflect such information. Standard classification methods developed outside astronomy, such as RF, SVM, or neural network classifiers, typical do not have a built-in option to account for the measurement error uncertainty.  On the other hand, classification methods originally developed for astronomical purposes may be equipped with a functionality to account for the measurement error uncertainty \citep[e.g.,][]{2011ApJ...729..141B, 2012ApJ...749...41B}. Since each perturbed data set does not contain the column for measurement error uncertainty, as shown in Figure~\ref{fig:data_after}, the latter classification methods cannot be used in the Gaussian perturbation framework.}

%In a simulation-based uncertainty quantification method like Gaussian perturbation, it is also possible to use a classifier that naturally takes into account the measurement error uncertainty \citep[e.g.,][]{2008ApJ...683...12B}.

\textcolor{black}{Choosing an appropriate classifier is one of the keys to an appropriate uncertainty quantification via the proposed Gaussian perturbation. Within the Gaussian perturbation framework, if we chose a classifier that naturally reflects measurement error uncertainty, this choice would cause an issue of using the same uncertainty information twice. This is because during the training the classifier will use the same information about measurement error uncertainty that has already been used to perturb the data. Consequently, the resulting classification uncertainty will be over-estimated in this case.}

% be the information about the measurement error is used once to perturb the data set, and then once again during the training by the built-in functionality of the chosen classifier.SClassifier twice On the other hand, we do not suggest adopting . 

\subsection{Correlated measurement error}\label{correlated}

In this work, we assume that measurement error in each feature is independent of other features because the measurements of $p$ features come with only $p$ measurement error uncertainties. This does \textit{not} mean that the correlation does not exist. In fact, modeling correlations among measurement errors is not unusual in astronomy, e.g., \textcolor{black}{\cite{kelly2007some} and} \cite{sereno2016bayesian} for linear regression. 

One particular difficulty arises in dealing with correlated measurement errors. The information about correlations among the $p$ measurement errors is necessary for constructing a full covariance structure of multiple measurement errors across features. But this information is typically not given in the data, which means that we need to \emph{estimate} all of the $p(p-1)/2$ correlations from the data. We note that these correlations are essentially the same as those among $p$ measurements (given the true values) because
\begin{align}
\textrm{Corr}(\xijobs,~ x^{\textrm{obs}}_{ij'})&=\textrm{Corr}(\xijtrue+\epsilon_{ij},~ x^{\textrm{true}}_{ij'}+\epsilon_{ij'})\nonumber\\
&=\textrm{Corr}(\epsilon_{ij},~ \epsilon_{ij'})\nonumber
\end{align}
for any $j\neq j'$. Therefore, we can estimate these correlations using the sample correlations in practice. 

There are possibly many ways to model these correlations within the Gaussian perturbation framework. A simple but naive approach is to construct the full $p\times p$ covariance matrix $V$ by filling out off-diagonal elements  with the estimated correlations. Then the resulting posterior predictive distribution of a perturbed observation is simply a multivariate version of \eqref{eq:ppd}, i.e.,
\begin{equation*}
x^{\textrm{rep}}_{i}\sim \mathcal{N}_p(x^{\textrm{obs}}_{i},~ 2V).
\end{equation*}
The notation $x^{\textrm{rep}}_{i}$ denotes $\{x^{\textrm{rep}}_{i1}, \ldots, x^{\textrm{rep}}_{ip}\}$ and, as defined before,  $x^{\textrm{obs}}_{i}=\{x^{\textrm{obs}}_{i1}, \ldots, x^{\textrm{obs}}_{ip}\}$.
One disadvantage of this approach is that it does not properly account for the uncertainty of estimating correlations. 

A Bayesian approach can be one alternative that enables modeling the correlations (or covariances) with their priors. Physical knowledge will be useful in constructing scientifically motivated priors, but flat priors between $-1$ and 1 would also be practical, \textcolor{black}{as done in \cite{sereno2016bayesian},} if such information were not available.  Then, the uncertainty of unknown correlations will be additionally reflected in the perturbed data sets. The resulting implementation, however, may become more computationally intensive as an inevitable cost for more elaborate Bayesian modeling. \textcolor{black}{Incorporating this Bayesian modeling approach to  measurement error correlations into any Bayesian classifiers, such as \cite{2011ApJ...729..141B, 2012ApJ...749...41B} or this work, might be an interesting future direction.}

\subsection{Gaussian perturbation for clustering}

The proposed Gaussian perturbation can naturally be extended to unsupervised learning problems, such as clustering analysis in astronomy. Extensive work has already been proposed relating to cluster stability evaluation via data perturbation, i.e., assessing how stable a clustering algorithm is to small perturbations in the data \citep{rand1971objective, breckenridge1989replicating, levine2001resampling, bhattacharjee2001classification, zhang2020cps}. In fact, simulating pseudo-data sets and ensembling the results have proven to be successful in achieving more robust clustering performance \citep{fridlyand2001applications, fern2003random, monti2003consensus, moeller2006performance}. Gaussian perturbation can serve as a novel perturbation scheme  especially suitable for assessing the variability in clustering results of astronomical clustering problems. This is because, as in the case of supervised learning, most standard clustering methods do not account for heteroscedastic measurement error uncertainties given in the astronomical data either. While accuracy metrics are not available in clustering tasks as in classification, the variability of any clustering metric can be similarly assessed, such as Silhouette Coefficient \citep{rousseeuw1987silhouettes}, Rand Index \citep{rand1971objective}, or any other measure of similarity or dispersion.

\subsection{Limitations}

Every statistical method has its own pros and cons, and the proposed method is no exception. The current work adopts a Gaussian measurement error model, but the Gaussian assumption may not always be sufficient to describe the complex nature of astronomical measurements well. For example, features such as color index at low signal-to-noise can have errors that are both non-Gaussian and asymmetrical \citep{babu2016skysurveys}. One may require different distributional assumptions depending on how the measurements were collected and how the  measurement error uncertainties were calculated. Therefore, it is desirable to extend the current work to encompass various measurement error models, such as a mixture of Gaussians and Student's $t$ measurement errors \citep{tak2019robust}. 

Also, the computational cost of Gaussian perturbation increases linearly in the number of simulations, which adds onto the  computational cost of a chosen classification method. In fact, the proposed approach increases the original computation cost of a classification method by a factor of $R$, the number of perturbed data sets. However, since any classification analysis can be independently conducted for each perturbed data set, the procedure is embarrassingly \textcolor{black}{parallelizable} over multiple cores. For example, the computational cost of the proposed approach can be restricted to that of a classification method, as long as 500 cores are available. Theoretically, the computational burden of the proposed approach can match that of a standard classification method if one has access to $R$ CPU cores.

\textcolor{black}{For instance, in the realistic data analysis for classifying high-$z$ quasars in Section~\ref{sec:astro}, it has taken 687 seconds in total to fit a random forest classifier with 50 trees to a single replicate data set. The implementation is conducted via \texttt{R} \citep{r2013} on a laptop equipped with 2.4GHz quad-core Intel Core i5 and 32 Gb RAM. Specifically, it takes 673 seconds to train the random forest classifier on 649,439  objects via 10-fold cross validation in the training set, and 14 seconds to predict labels of 1,862,968 objects in the prediction set. The total computational time to fit the random forest to all of the 500 replicate data sets is about 351,000 seconds (about 4 days) without parallelization, but is about 700 seconds with parallelization over 500 cores.}
%(We note that using a virtual parallization package, such as \texttt{doParallel} in \texttt{R}, which enables running more parallel jobs  than the actual number of cores, may increase the average computational cost of a single fit.)

% We note that using a virtual parallization package, such as \texttt{doParallel} in \texttt{R}, which enables running more parallel jobs  than the actual number of cores, may increase the average computational cost of a single fit.}

%in the prediction set 
%  The implementation is conducted on the ROAR system, one of the high performance computation systems managed by the Institute for Computational and Data Sciences at the Pennsylvania State University. The specification of thA virtual parallelization \texttt{R} package, \texttt{doParallel}
% A virtual parallelization \texttt{R} package, \texttt{doParallel}, is used to implement the pr on virtual 500 cores.

%%%%%%%%%%%%%%%%%%%%%%%%%%%%%%%%%%%%%%%%%%%%%%%%%%%%%%%%%%%%%%
\section{Concluding Remarks} \label{sec:conclusion}
%%%%%%%%%%%%%%%%%%%%%%%%%%%%%%%%%%%%%%%%%%%%%%%%%%%%%%%%%%%%%%

Astronomical data are unusual in the sense that each measurement is accompanied by heteroscedastic measurement error whose uncertainty is known. These uncertainties are often ignored in astronomical classification problems because standard classification methods, such as support-vector machine and random forest, cannot incorporate them. This work proposes Gaussian perturbation as a simulation-based  way to incorporate the measurement error uncertainties into any standard classification methods to better quantify classification uncertainty. The key idea is to simulate pseudo-data sets from a posterior predictive distribution of a Gaussian measurement error model, using the known heteroscedastic measurement error uncertainty.  Then, any chosen standard classification method can be fit on each of these simulations. The resulting variation of a quantity of interest across the multiple fits naturally reflects the measurement error uncertainty as it has been propagated through every step of the  procedure. We have illustrated this procedure via an extensive simulation study using SVM  and random forest. Additionally, we have demonstrated its potential for astronomical applications through the problem of classifying high-$z$ quasars from astronomical catalog data.

We note that this is not the only work raising a question about how to incorporate the unusual feature of astronomical data into astronomical data analyses, as described in the introduction.  More recently, a small group of astronomers and one of the co-authors of this work had an active discussion about how to  incorporate measurement error in astronomical data analyses during a three-day workshop, Petabytes to Science\footnote{\url{https://petabytestoscience.github.io/workshop-iii}}, held in Boston in Nov, 2019. Later, one of the workshop organizers  carefully examined how the noises of input data propagate to a result of deep learning regression  via an analytically tractable  single pendulum experiment \citep{caldeira2020}. We hope that this work adds a momentum for the community to continue a discussion about incorporating measurement error in various contexts of astronomical data analyses.

\acknowledgments

SS and HT appreciate the Pennsylvania State University’s Institute for Computational and Data Sciences for its computational support via the Roar supercomputer. HT thanks the Kavli Foundation and AURA for  travel support to the workshop, Petabytes to Science, held in Boston in 2019. JDT appreciates  support from NASA
ADP grant 80NSSC18K0878, Chandra X-ray Center grant GO0-21080X, the V.~M.~Willaman Endowment, and Penn State ACIS Instrument Team Contract SV4-74018 (issued by the Chandra X-ray
Center, which is operated by the Smithsonian Astrophysical Observatory for and on behalf of NASA under contract NAS8-03060).  We thank Jackeline Moreno and Weixiang Yu for their helpful discussions about this work.
%\textcolor{blue}{GJB appreciates ...}

%\software{R}

%\appendix

%\newpage


\bibliography{main}
\bibliographystyle{aasjournal}



\end{document}